\documentclass[12pt, a4paper]{article}

\usepackage[utf8]{inputenc}
\usepackage[T1]{fontenc}
\usepackage{graphicx}
\usepackage{amsmath}
\usepackage{hyperref}
\usepackage{float}
\usepackage{color}
\usepackage{subcaption}
\usepackage{multirow}

\usepackage{authblk}

\usepackage[numbers]{natbib}  

\title{Characterizing Quantum Internet Using Complex Network Models}

\author[1,2]{Otávio José R. Silveira}
\author[2]{Nycolas B. da Silva}
\author[3]{Saulo L. L. da Silva}
\author[1]{Angélica Sousa da Mata\thanks{angelica.mata@ufla.br}}

\affil[1]{Departamento de F\'{\i}sica, Instituto de Ciências Naturais, Universidade Federal de Lavras, Caixa postal 3037, CEP: 37200-900, Lavras, MG, Brazil}
\affil[2]{Instituto de Física Gleb Wataghin, UNICAMP, Rua Sérgio Buarque de Holanda, 777, Cidade Universitária, CEP: 13083-859, Campinas, SP, Brazil}
\affil[3]{Centro Federal de Educação Tecnológica de Minas Gerais, Avenida Monsenhor Luiz de \\ Gonzaga, 103, CEP: 37250-000, Nepomuceno, MG, Brazil}
\date{\today}

\begin{document}

\maketitle

\begin{abstract}
Quantum communication is a growing area of research, with quantum internet being one of the most promising applications. Studying the statistical properties of this network is essential to understanding its connectivity and the efficiency of the entanglement distribution. However, the models proposed in the literature often assume homogeneous distributions in the connections of the optical fiber infrastructure, without considering the heterogeneity of the network. In this work, we propose new models for the quantum internet that incorporate this heterogeneity of node connections in the optical fiber network, analyzing how this characteristic influences fundamental metrics such as the degree distribution, the average clustering coefficient, the average shortest path and assortativity. Our results indicate that, compared to homogeneous models, heterogeneous networks 
efficiently reproduce key structural properties of real optical fiber networks, including degree distribution, assortativity, and hierarchical behavior. These findings highlight the impact of network structure on quantum communication and can contribute to more realistic modeling of quantum internet infrastructure.\end{abstract} 

\section{Introduction}\label{intro}

In recent years, quantum technologies have advanced significantly, opening up new possibilities for computing and communication. Quantum computing, in particular, represents a new computing paradigm that uses the properties of quantum systems to solve problems that are impractical for classical computers \cite{nielsen2010quantum}. Quantum algorithms, such as Shor's for factoring integers \cite{shor1999polynomial} and Grover's for searching unstructured data \cite{grover1996fast}, demonstrate significant advantages over classical methods, suggesting potential impacts in areas such as cryptography \cite{gisin2002quantum,scarani2009security}, optimization \cite{harrow2009quantum} and simulation of physical systems \cite{quantum_simulation1,quantum_simulation2}. Although current devices are still in their early stages, the development of quantum computing significantly impacts important areas for technological development. As quantum computing grows, it becomes essential to develop infrastructures to enable secure communication between quantum devices, which leads to the need for advances in quantum communication and, consequently, the quantum network.

Quantum communication faces significant challenges due to the fragility of quantum states. Decoherence and loss of information along quantum channels limit the distance and reliability of qubit transmission. To overcome these difficulties, solutions have been developed such as quantum repeaters \cite{repeater0,repeater1,repeater2}, which allow entanglement to be distributed over long distances, quantum memories \cite{memories0,memories1,memories2}, used to store quantum states for a coherent period of time, and quantum error correction protocols, which protect information from undesired noise \cite{quantum_error}. By overcoming these challenges, along with the engineering obstacles associated with integrating different quantum technologies, a future quantum internet will enable the secure and efficient transmission of quantum information between geographically dispersed devices \cite{quantum_internet1,quantum_internet2}. In addition, the construction of this network can benefit from the existing infrastructure of the classical internet, using fiber optic networks and other components to facilitate the implementation and scalability of quantum communication.

In this sense, some studies recognize the importance of entanglement distribution as a fundamental element for building a quantum internet network \cite{distribution1,distribution2,distribution3,distribution4}. These studies explore various methods for generating and sharing pairs of entangled qubits between distant nodes and propose algorithms and protocols for optimizing the distribution of entanglement along different network paths.
On the other hand, some works start from the assumption that the technologies and methods for distributing entanglement are already well established and consider that an underlying structure of a quantum network already exists, focusing on the statistical properties of the quantum internet \cite{statistical_net1,statistical_net2,statistical_net3}. In particular, the reference \cite{brito2020statistical} proposes a quantum internet model based on an optical fiber infrastructure, showing that these networks, like random graphs, have a phase transition between a weakly connected state and a fully connected state. However, these networks do not have the small-world property, which can limit the efficiency of entanglement distribution. In \cite{brito2021satellite}, the authors demonstrate that, when considering a network made up of satellites, this property emerges, indicating a potential increase in the efficiency of entanglement distribution. In addition, other studies, such as \cite{oliveira2024statistical}, have examined how a non-uniform spatial distribution of nodes in the optical fiber network affects its statistical properties, under the assumption that node connectivity remains homogeneous, particularly regarding its influence on entanglement distribution.

Our work, inspired by the model proposed in Ref.~\cite{brito2020statistical}, investigates new approaches for modeling the structure of the quantum internet by incorporating heterogeneity into the links fiber optic networks. It is important to note that these are adapted models; rather, they are models already established in the literature, applied here for the first time in the context of quantum internet networks. This approach takes into account that real networks are not homogeneous, but have heterogeneity in the distribution of their connections influenced by economic, technological, and geopolitical factors in the region where they are located, which affect the distribution and connectivity of nodes. Based on these models, we construct the fiber-optic networks that serve as the physical infrastructure on which the quantum (photonic) networks operate. The statistical properties of these photonic networks are then analyzed and compared with those of the homogeneous network proposed by Brito et al. \cite{brito2020statistical}. Additionally, we compare our proposed models with the one used in Ref.~\cite{brito2020statistical} for real fiber optic networks from six different continents, highlighting the importance of taking into account the heterogeneity of such structures. 

Our results show that heterogeneous network models achieve comparable efficiency to the denser homogeneous Brito et al. model, but with fewer connections, as high-degree nodes act as shortcuts linking distant regions of the network. These models reproduce key structural properties observed in photonic networks running on top of real optical fiber networks, including the degree distribution, slightly disassortative behavior, and a power-law decay of the clustering coefficient as a function of degree. Such features are characteristic of hierarchical networks and have also been observed in real internet structures \cite{internet_power_law}. Comparisons across six continents indicate strong qualitative agreement between the heterogeneous models and real networks, highlighting that these models can provide a realistic representation of the emerging quantum internet.

The paper is organized as follows: In Section~\ref{sec:methodology}, we briefly present our methodology. A more detailed explanation can be found in the Supplementary Material (SM), where the main metrics used to characterize networks are introduced in Section~S1, the model proposed by Brito et al.~\cite{brito2020statistical} is revisited, and two new models are presented in Section~S2. In Section~\ref{sec:results}, we present our results, discussing the statistical properties of the newly proposed networks and comparing them with real networks. In Section~S4 of the SM, additional results are reported, including for real networks. Finally, in Section~\ref{sec:conclusions}, we draw our conclusions.

\section{Methodology}
\label{sec:methodology}
\subsection{Networks: Definitions and Metrics}\label{subsec:network_fundamental}

A network consists of $N$ nodes, representing the components of the system, connected by links. In the context of quantum networks, these nodes can be quantum circuits, quantum memories, or quantum repeaters~\cite{memories1,repeater1}, which are connected through quantum channels such as optical fibers~\cite{channel}. To characterize the structure of these photonic networks operating on the fiber-optic infrastructure, we employed several fundamental metrics from complex network theory. The average shortest path length $\langle l \rangle$ indicates the average number of links along the shortest paths connecting all pairs of nodes. The clustering coefficient $\langle C \rangle$ quantifies the tendency of nodes to form tightly connected groups, while the clustering coefficient as a function of node degree $C(k)$ evaluates how local clustering varies with connectivity. Assortativity $r$ indicates the tendency of nodes to connect with other nodes of similar degree, and the average degree of nearest neighbors as a function of node degree $k_{\text{nn}}(k)$ captures the connectivity of a node's neighbors relative to its own degree~\cite{network_barabasi,Mata2020}. A full description of these metrics is provided in Section~S1 of the Supplementary Material (SM).

\subsection{Models for Quantum Networks}\label{subsec:models}

In the context of the quantum internet, network theory provides a fundamental mathematical framework for studying the statistical properties of this system. In \cite{brito2020statistical}, the authors proposed a model that assumes the presence of a conventional internet infrastructure connected by optical fibers. They used Waxman’s graph \cite{waxman1988routing} to represent the connections, where each fiber links two nodes in the network. A key characteristic of this model is that its connectivity distribution follows a Poisson distribution, resembling that of random graphs. Consequently, most nodes have a similar number of connections, leading to a relatively homogeneous network. With this structure, pairs of entangled photons can be established between the nodes through the fibers, considering a probability associated with the transmission of the photons. 
The construction of the model follows the steps described below:

\begin{enumerate}
    \item $N$ nodes are distributed uniformly on a disk of radius $R$, whose area must correspond to the region where the network will be implemented. For example, for the USA, $R = 1800 \text{ km}$ is used. \label{first_step}
    \item An optical fiber network is built by connecting nodes based on Waxman's model. Two nodes $i$ and $j$ are connected with a probability given by:
    
    $$\Gamma_{ij}=\beta e^{-d_{ij}/\alpha L},$$
    
    where $d_{ij}$ (km) is the Euclidean distance between the nodes, $L$ is the maximum distance in the network, $\alpha$ regulates the typical length of the connections, and $\beta$ defines the degree of connectivity. Here is used $\alpha L = 226 \text{ km}$ and $\beta=1$. \label{second_step}
    \item After establishing the optical fiber infrastructure, the photonic network is constructed, it means, it is evaluated whether the connections can share pairs of entangled photons. Photonic losses along the fibers are considered, determined by the transmissivity

    \begin{equation}\label{eq:transmissivity}
        q_{ij} = 10^{-\gamma d_{ij}/10},
    \end{equation}
    where $\gamma = 0.2 \text{ dB/km}$ is the fiber loss coefficient for silicon. As the fiber distance increases, the photonic losses increase exponentially. The probability of entanglement between two nodes is given by:
    
    $$p_{ij}=1-(1-q_{ij})^{n_p},$$
    
    where $n_p$ is the number of photons transmitted per node. In this model, $n_p  = 1000$ is assumed. It should be noted that this probability is based solely on transmissivity and does not include the effects of quantum devices such as repeaters or quantum memories. \label{third_step}
\end{enumerate}

Based on the model proposed by Brito et al.~\cite{brito2020statistical}, this work explores two other models for the quantum internet, modifying step \ref{second_step} of the construction process. The main objective is to investigate how characteristics that make the optical fiber infrastructure more heterogeneous affect the connectivity and phase transition of the quantum network. The motivation for this approach arises from the fact that homogeneous networks are unlikely in practice. Fiber optic network infrastructure generally shows high heterogeneity, influenced by economic factors, population density, and technological capabilities \cite{network_barabasi,caida_skitter}.

The main feature of the first model proposed is the growth with a preferential attachment. Each node in the network is assigned a random position according to the step \ref{first_step}, but it is only actually positioned during the growth process.
Growth is based on the model proposed by Soares et al. \cite{soares2005preferential}. and the fiber optic connections described in the step \ref{second_step} are made based on the following preferred connection rule:

\begin{enumerate}
    \setcounter{enumi}{1}
    \item Initially, the first $m+1$ nodes are positioned and connected to each other. Then, each incoming node is added to the network and connects to $m$ existing nodes with a probability that depends on the number of connections $(k_i)$ and the distance $(d_{ij})$ between the existing node $i$ and the incoming node $j$. The preferential connection probability is given by:
    
    $$\Lambda_{ij} = \frac{k_id_{ij}^{-\alpha_A}}{\sum_{i=1}^{n}k_id_{ij}^{-\alpha_A}},$$
    
    where $n$ is the number of nodes in the network under formation and $\alpha_A$ is a model parameter that adjusts the typical distance of the connections.
\end{enumerate}

At the end of the process, a heterogeneous fiber optic network is formed, in which connections are established not only preferentially with highly connected nodes but also influenced by the nodes’ spatial proximity. It is important to note that this model produces a degree distribution that follows a q-exponential fit, characterizing the network as heterogeneous for small values of $\alpha_A$, whereas it approaches a homogeneous structure as $\alpha_A$ tends to infinity \cite{soares2005preferential,piva2021networks}. Next, the construction of the photonic network, it means, the distribution of entanglement in the fiber optic network follows step \ref{third_step}.

The second model proposed in this work is characterized by the formation of a scale-free optical fiber network. Initially, the nodes are randomly distributed in the plane, as described in step \ref{first_step}. The connection step for forming the fiber optic network (step \ref{second_step}) is then modified based on the Rozenfeld et al model \cite{rozenfeld2002scale}, as follows:

\begin{enumerate}
    \setcounter{enumi}{1}
    \item Each node is associated with a number $k$ of links obtained from the scale-free distribution:
    
    \begin{equation}\label{eq:scale-free-distribution}
        \Pi(k)=Ck^{-\lambda}, \qquad m < k <K,
    \end{equation}
        
    where $m$ is the minimum degree, $K$ is the maximum degree and $\lambda > 2$ controls the heterogeneity of the network, determining how the node degrees are distributed. The normalization constant is $C \approx (\lambda - 1)m^{\lambda - 1}$ for large $K$. After defining $k$ for each node, a node is chosen at random and connected to its nearest neighbors until it reaches its assigned connectivity $k$ or until it explores all nodes within a radius:
    
    \begin{equation}\label{eq:scale-free-radius}
        r(k)=A\sqrt{k},
    \end{equation}
    
    where $A$ is a model parameter that defines the scale of the maximum distance at which connections can be made. In this work, we chose to keep $A=100 \text{ km}$, a typical distance for connections within the disk of radius $R=1800 \text{ km}$. Additionally, it's being consider $m=3$ and $K=1 \times 10^{6}$ as fixed models parameters. It is important to note that not all desired connections can be established, as neighboring nodes also have a limit $k$ which, once reached, prevents new connections.
\end{enumerate}

Next, the step \ref{third_step} is done to distribute the entanglement in the network. This mechanism for forming optical fibers results in a highly heterogeneous network, with few highly connected nodes and many nodes with few connections, a typical characteristic of scale-free networks. Section S2 of the SM provides samples of each model and the key statistical properties of the Brito et al. model, while the statistical properties of the new models are presented in the Results section.

To validate our approach, we also analyze real internet topologies obtained from the August 2020 Internet Topology Data Kit (ITDK) maintained by CAIDA\footnote{\url{https://www.caida.org/}}. Specifically, we consider backbone-level networks for six continents: North America, South America, Europe, Africa, Oceania, and Asia. These empirical infrastructure networks allow us to compare the structural properties of real-world communication systems with those generated by our models, highlighting both similarities and differences. The technical details regarding data acquisition and preprocessing are provided in the section S3 of the SM.

\section{Results}
\label{sec:results}
\subsection{Properties of Quantum Network Models}


This section presents the main statistical properties of the new models for the quantum internet. Numerical simulations were carried out with 100 samples for each model. In cases where the simulations are more complex and time-consuming, a smaller number of samples was used, which will be indicated in the text. For convenience, the model based on Soares et al. \cite{soares2005preferential} will be referred to as the Brito-Soares model, while the model based on Rozenfeld et al. \cite{rozenfeld2002scale} will be referred to as the Brito-Rozenfeld model. This nomenclature was chosen because we adapted both models according to Brito's preliminary assumption that the nodes are uniformly distributed within a disk of radius $R$, with an area corresponding to the region where the network will be implemented. Beyond the homogeneous distribution of nodes, the new models also preserve Brito’s original considerations regarding the formation of the quantum network, which constitute a more fundamental aspect of their design.

\begin{figure}[H]
    \centering
    \includegraphics[width=0.9\linewidth]{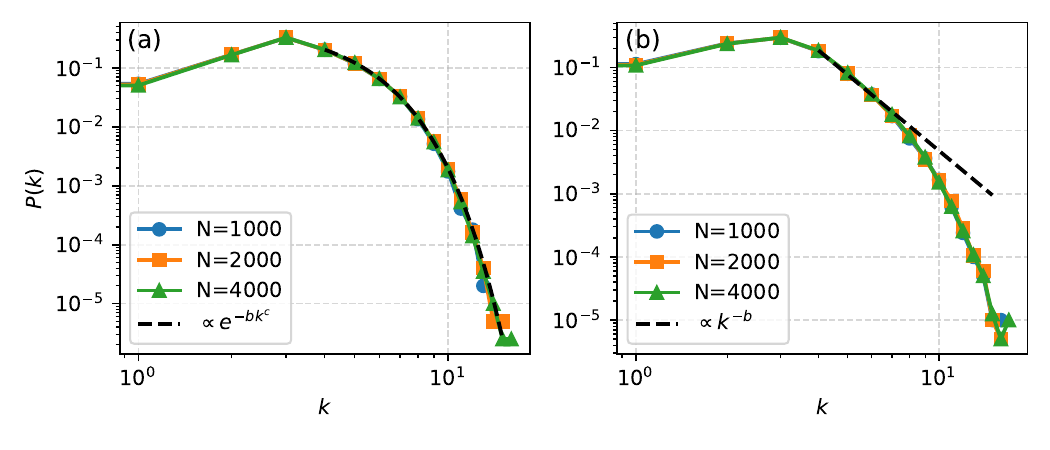}
    \caption{Degree distribution for networks with different numbers of nodes and density $\rho = 8 \times 10^{-5}$. (a) Brito-Soares model. (b) Brito-Rozenfeld model. The distributions collapse into a single curve  for different values of N in both models. The dashed black curve represents the model-specific fit to the tail, using either an exponential or a power-law fit as appropriate. The fitting parameters are given in the text.}
    \label{fig:degree_distrib_for_N}
\end{figure}

The first property analyzed is the degree distribution of the networks. As discussed earlier, the models were developed to capture the heterogeneity of the optical fiber network, whose degree distribution can follow a power law, since most real networks behave like that, including technological ones \cite{caida_skitter,barabasi2009scale}. However, the photonic network should not follow exactly the same distribution, as not all optical connections share entangled photon pairs. Figure \ref{fig:degree_distrib_for_N} shows the degree distributions of the photonic networks occurring on optical fiber networks generated by (a) the Brito-Soares model and (b) the Brito-Rozenfeld model, as a function of the total number of nodes for a fixed density $\rho = N / \pi R^2$.

Figure \ref{fig:degree_distrib_for_N} shows that, for both models, the degree distributions collapse into a single curve for different values of $N$.
The presence of a long tail suggests typical power-law behavior. This pattern indicates that most nodes of the network have few connections, but the tail of the distribution reveals the existence of a few highly connected nodes, which gives the network a certain heterogeneity. 

For the Brito-Soares model, shown in Figure \ref{fig:degree_distrib_for_N}(a), the data were fitted with a generalized exponential distribution, $P(k) = a\exp^{-bk^{c}}$. The parameters found were $a=0.5$, $b=0.065$ and $c=1.94$. For the Brito-Rozenfeld model, in Figure \ref{fig:degree_distrib_for_N}(b), the tails of the distributions were fitted by a power law, $P(k) = ak^{-b}$, with $a=47.5$ and $b=4.0$. It is worth noting that, in the case of the Brito-Rozenfeld model, the exponent of the degree distribution of the optical fiber network $\Pi(k)=Ck^{-\lambda}$, was $\lambda = 3$. The fit found was slightly different, which suggests that the heterogeneity of the network contributes to this value, but with a slightly higher exponent, since not all fiber optic connections share entangled photons, accentuating the tail of the distribution observed in Figure \ref{fig:degree_distrib_for_N}(b). However, both exhibit the same behavior in the degree distribution of the optical fiber networks generated by their respective underlying models, as discussed in Section \ref{subsec:models}.

The degree distribution for different node densities is shown in Figure S5 of the SM, where we observe a clear dependence of the distribution on the density. By keeping the total number of nodes $N$ fixed and increasing the density, the radius $R$ of the disk in which the nodes are placed is reduced. Consequently, the nodes become closer to each other, and according to the preferential attachment rule in both models, this increases the likelihood of forming more connections.

\begin{figure}[H]
    \centering
    \includegraphics[width=0.9\linewidth]{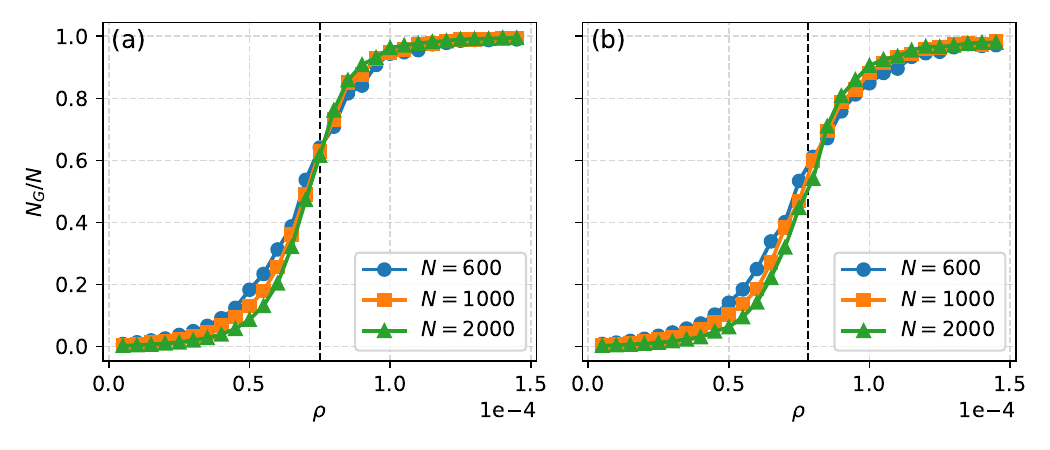}
    \caption{Phase transition as a function of $\rho$, for different numbers of nodes. The black dashed curve indicates the critical density of the phase transition. (a) Brito-Soares model ($\rho_c \approx 7.5 \times 10^{-5}$) and (b) Brito-Rozenfeld model ($\rho_c \approx 7.8 \times 10^{-5}$).}
    \label{fig:phase_transition_for_rho}
\end{figure}

As shown in Ref.\cite{brito2020statistical}, the Brito model exhibits a phase transition between a disconnected state, characterized by a few isolated clusters, and a highly connected state, where a giant component emerges that interconnects most of the network, as illustrated in Figures S3  and S4 of the SM. To analyze this behavior, we calculate the fraction of nodes in the giant component, $N_G/N$, as a function of node density. The critical point of the phase transition is identified by the peak in the standard deviation of $N_G/N$, which signals the onset of large-scale connectivity fluctuations. Figure \ref{fig:phase_transition_for_rho} shows this transition for our two new models. A second-order phase transition is observed, similar to that described in Ref.\cite{brito2020statistical}, as well as a universal behavior of the transition as a function of density. For the Brito-Soares model, the critical density is $\rho_c \approx 7.5 \times 10^{-5}$, while for the Brito-Rozenfeld model, $\rho_c \approx 7.8 \times 10^{-5}$. These values are higher than those found in Ref.\cite{brito2020statistical}, where $\rho_c \approx 6.82 \times 10^{-5}$. This occurs because the new models exhibit greater heterogeneity, requiring a higher node density to ensure network connectivity and the formation of the giant component. Figure S6 of the SM illustrates this phase transition for different values of $R$, allowing us to estimate the minimum number of nodes required to form a highly connected network within a given area for an arbitrary $R$.

\begin{figure}[H]
    \centering
    \includegraphics[width=1\linewidth]{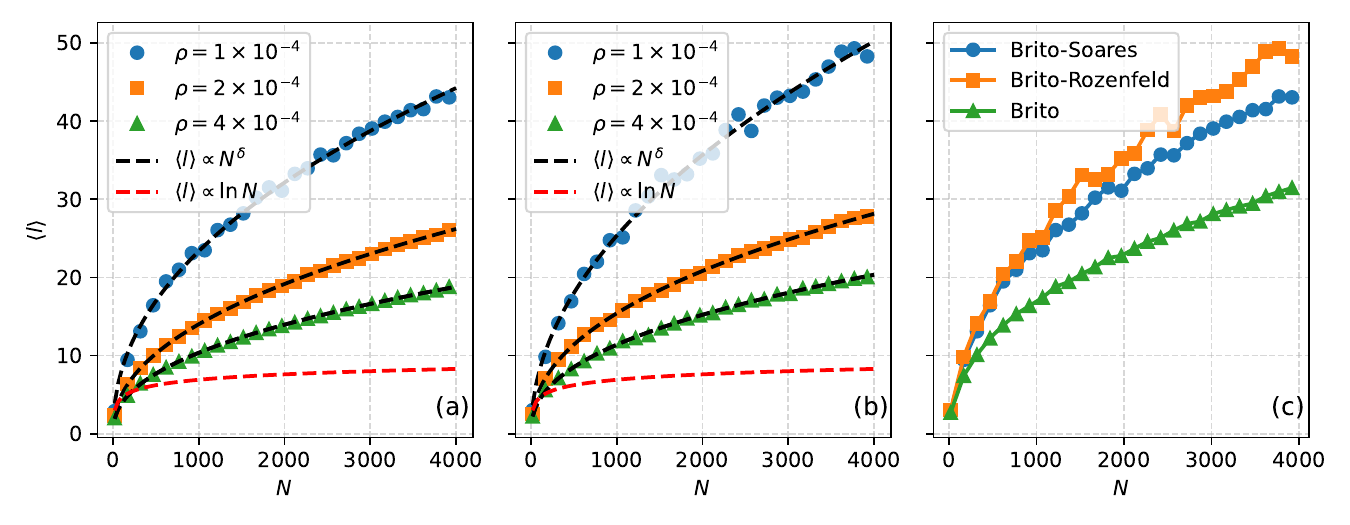}
    \caption{Average shortest path length as a function of the number of nodes. (a) Brito-Soares model. (b) Brito-Rozenfeld model. The dashed curves indicate fits for $\left \langle l \right \rangle = a N^{\delta} / \rho$ (black) and $\left \langle l \right \rangle \propto \ln N$ (red), showing that the networks do not have the small-world property. (c) Comparison of the average shortest path length between the three models studied, for $\rho = 1 \times 10^{-4}$.}
    \label{fig:av_spl}
\end{figure}

The average shortest path length $\left \langle l \right \rangle$ of the networks is shown in Figure \ref{fig:av_spl}, where the simulations were carried out with 25 samples. It can be seen that the photonic networks analyzed follow the same behavior as the Brito et al. model \cite{brito2020statistical}, i.e. they do not exhibit the small-world property, for which $\left \langle l \right \rangle \propto \ln N$. Instead, the data are better fitted by $\left \langle l \right \rangle \sim N^{\delta} / \rho$. For the traditional Brito model, we obtain $\delta \approx 0.45$, while for the Brito-Soares model, $\delta$ takes the values $0.46$, $0.45$, and $0.43$ for increasing densities (shown in Figure \ref{fig:av_spl}). For the Brito-Rozenfeld model, $\delta$ is $0.49$, $0.43$, and $0.41$, respectively. As expected, the average shortest path decreases with the increase of the density, because as the network becomes denser, the nodes become more connected, decreasing the effective distance (number of links) between them.

\begin{figure}[H]
    \centering
    \includegraphics[width=0.7\linewidth]{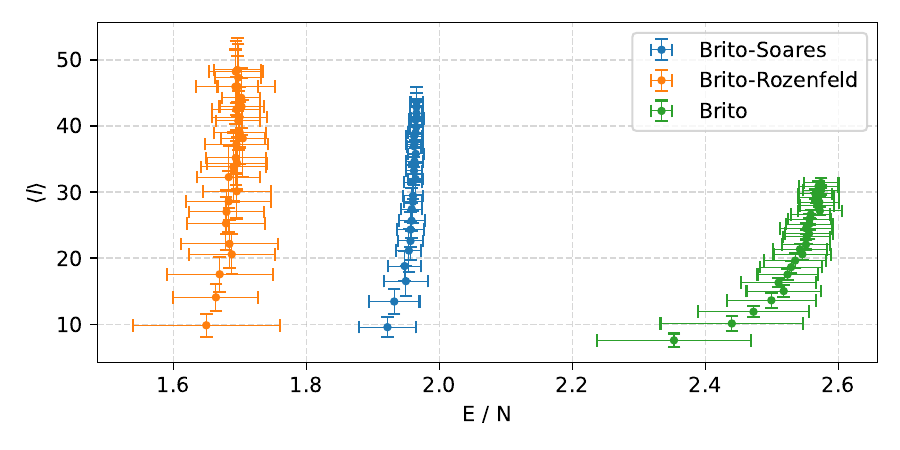}
    \caption{Comparison of average shortest path length as a function of the normalized number of edges ($E/N$) for the different network models. The Brito et al. model requires a higher number of connections to achieve a given value of $\left \langle l \right \rangle$.}
    \label{fig:av_spl_weighted}
\end{figure}

Figure \ref{fig:av_spl}(c) also shows a comparison of the average shortest path length between the three models studied. It can be seen that, for the same number of nodes, the Brito et al. model has the lowest $\left \langle l \right \rangle$ values, followed by the Brito-Soares model and, lastly, the Brito-Rozenfeld model.
At first, this result may seem contradictory, since models with heterogeneous degree distributions are expected to have smaller average shortest path lengths than homogeneous ones, due to the presence of hubs that act as shortcuts between distant parts of the network. However, the average shortest path length observed in the Brito et al. model is not a result of heterogeneity, but rather of its higher link density. For the same number of nodes, this model generates significantly more edges than the others, increasing connectivity and reducing the average shortest path, as seen in Figures S1, S3 and S4 of the SM. Figure \ref{fig:av_spl_weighted} shows the average shortest path length as a function of $E/N$, the number of edges normalized by the number of nodes. Each point in the plot corresponds to a single network, sampled for different values of $N$. The figure shows that, to reach a given average shortest path length, the Brito et al. model requires a much higher $E/N$ compared to the others. This indicates that the other models, thanks to their heterogeneous structure, achieve similar efficiency with fewer connections, as high-degree nodes act as shortcuts linking distant regions of the network.

Furthermore, as discussed in Ref.\cite{brito2020statistical}, the average shortest path of photonic networks is directly related to the number of quantum repeaters needed for the distribution of entangled photons. Considering a network with $R=1800$~km and $N=1000$, we have a density of $\rho \approx 1 \times 10^{-4}$. In this scenario, the average shortest path length is approximately $\left \langle l \right \rangle \approx 17$ for the Brito et al. model, $\left \langle l \right \rangle \approx 24$ for the Brito-Soares model, and $\left \langle l \right \rangle \approx 25$ for the Brito-Rozenfeld model.

\begin{figure}[H]
    \centering
    \includegraphics[width=0.8\linewidth]{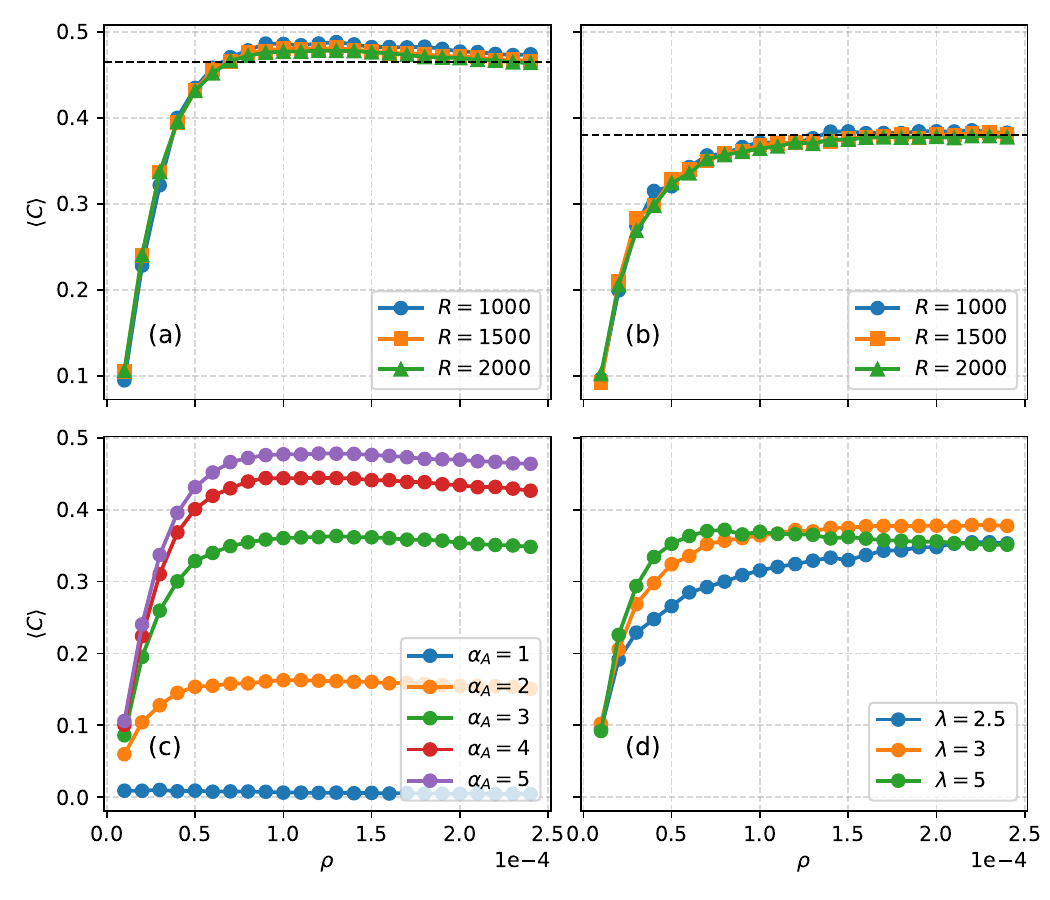}
    \caption{Average clustering coefficient as a function of $\rho$ for different $R$. (a) Brito-Soares model with $\alpha_A=5$. (b) Brito-Rozenfeld model with $\lambda=3$. In both, the curves collapse into a single curve with asymptotic behavior, with asymptotic values indicated in the text. (c) and (d) show $\left \langle C \right \rangle$ for different model parameters, with $R=2000$.}
    \label{fig:av_cluster}
\end{figure}

The average clustering coefficient $\left \langle C \right \rangle$ as a function of $\rho$ is shown in Figure \ref{fig:av_cluster}. The top panels show that all the curves collapse into a single behavior, suggesting universality and an asymptotic value as the number of nodes increases. For the Brito-Soares model, the asymptotic value $\left \langle C \right \rangle \approx 0.47$, indicates that the photonic network, like that of Brito et al., is extremely aggregated.
Similarly, the Brito-Rozenfeld network shows $\left \langle C \right \rangle \approx 0.38$. Although this value is lower, the network still maintains high aggregation, as also illustrated in Figure S4 of the SM.

It is also interesting to observe the behavior of $\left \langle C\right \rangle$ as a function of the model parameters, as can be seen in the bottom panels of Figure \ref{fig:av_cluster}. In the Brito-Soares model, the parameter $\alpha_A$ directly affects the asymptotic behavior of $\left \langle C \right \rangle$, which increases with $\alpha_A$. This occurs because the preferential attachment probability ($\Lambda_{ij}$) decreases with the distance between nodes $i$ and $j$, but smaller values of $\alpha_A$ weaken this dependence, allowing more distant nodes to connect, i.e., $\Lambda_{ij} \sim d_{ij}^{-\alpha_A}$. As the transmission of entangled photons suffers exponential attenuation as given by  $q_{ij} = 10^{-\gamma d_{ij}/10}$, excessively long connections become unfeasible, resulting in poorly connected networks with few clusters.
In contrast, the $\lambda$ parameter of the Brito-Rozenfeld model does not significantly impact the asymptotic value of $\left \langle C \right \rangle$, but only changes the initial shape of the curve, for small values of $\rho$.

\begin{figure}[H]
    \centering
    \includegraphics[width=1\linewidth]{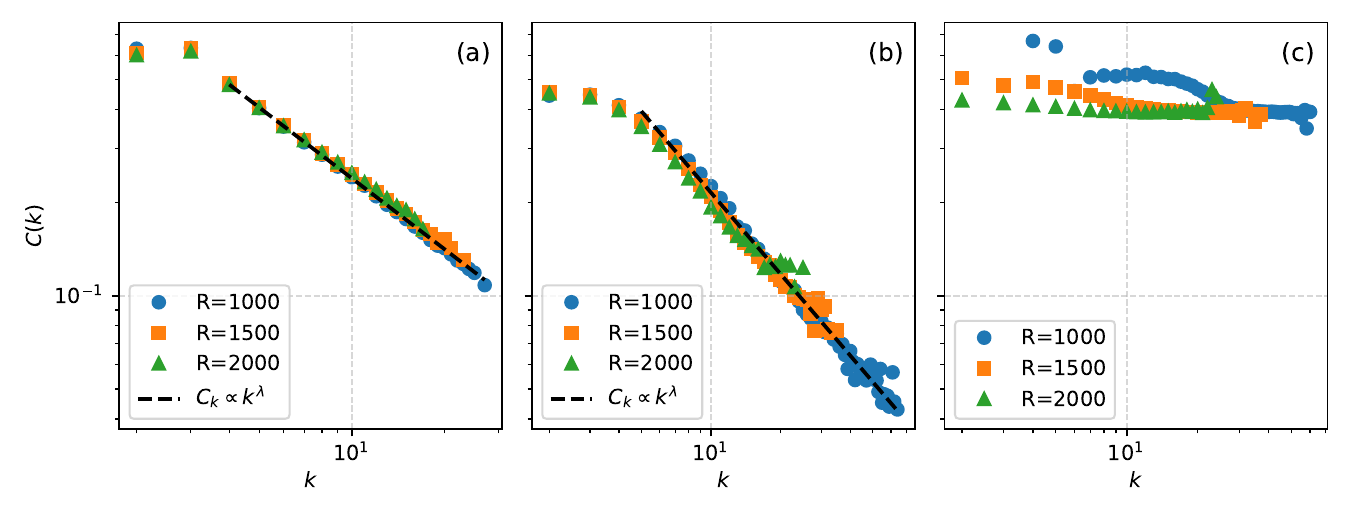}
    \caption{Average clustering coefficient as a function of $k$ for the three network models. For the Brito-Soares (a) and Brito-Rozenfeld (b) models, a power-law behavior is observed, which is characteristic of hierarchical networks. In contrast, this behavior is not present in the Brito et al. model (c).}
    \label{fig:cluster_k}
\end{figure}

Another interesting property to analyze is the clustering coefficient as a function of node degree, $C(k)$. Figure \ref{fig:cluster_k} shows $C(k)$ for the photonic networks generated by the different models, with $N = 2000$. For the Brito et al. model, $C(k)$ remains approximately constant, meaning that low-degree and high-degree nodes cluster in a similar way. In contrast, the Brito-Soares and Brito-Rozenfeld models exhibit a power-law behavior, where the dashed black curve corresponds to the fit $C(k) \propto k^{\lambda}$. The fitted exponents are $\lambda = -0.76$ for the Brito-Soares model and $\lambda = -0.87$ for the Brito-Rozenfeld model. This type of behavior is typical of hierarchical networks, where distinct groups of nodes are loosely connected to each other, forming a layered structure of connectivity \cite{barabasi_hierarchical}. It is also worth noting that such behavior is observed in the current structure of the internet, indicating that, in this aspect, the models proposed here are more accurate than the one introduced by Brito et al. \cite{internet_power_law}.
All analyses shown in Figures \ref{fig:degree_distrib_for_N}, \ref{fig:phase_transition_for_rho}, \ref{fig:av_spl} and \ref{fig:av_cluster} for the Brito-Rozenfeld and Brito-Soares models can also be found for the original Brito et al. model in Figure S2 of the SM.

\begin{figure}[H]
    \centering
    \includegraphics[width=1\linewidth]{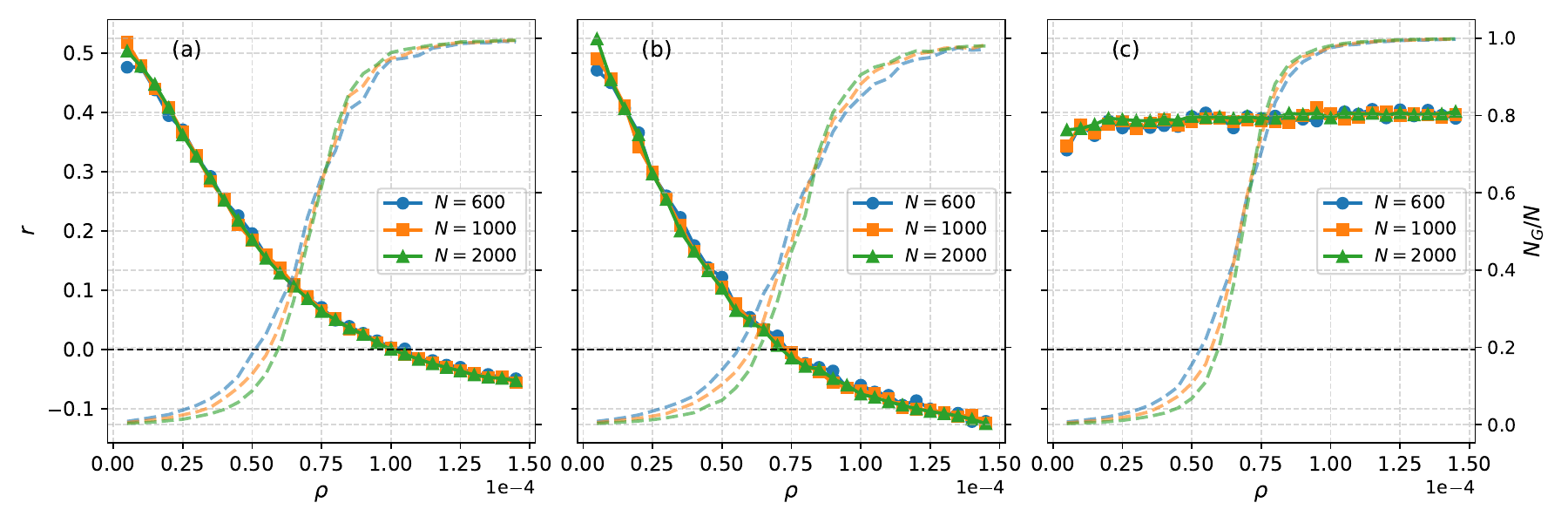}
    \caption{Assortativity and phase transition as a function of $\rho$ for different $N$. The solid curves with symbols represent the assortativity (left axis - $r$) and the dashed faded curves indicate the phase transition (right axis - $N_G / N$). The dashed black line marks $r=0$, separating assortative networks ($r>0$) from disassortative ones ($r<0$). (a) Brito-Soares model. (b) Brito-Rozenfeld model. (c) Brito et al. model.}
    \label{fig:assortativity}
\end{figure}

The assortativity $r$ of quantum network models is analyzed in Figure \ref{fig:assortativity}, which shows the variation of this parameter as a function of node density $\rho$ for various $N$. It can be seen that all the curves collapse into a single curve, suggesting a universal behavior for $r$ as a function of $\rho$. 
 $r = 0$ indicates no correlation, while $r > 0$ and $r < 0$ correspond to assortative and disassortative networks, respectively.
 In the Brito-Soares and Brito-Rozenfeld models, the networks are assortative for small values of $\rho$, i.e. the nodes tend to connect to others with similar connectivity degrees. This is because, in these regimes, the network is not yet completely connected, so there are just a few connected nodes in the network. This effect can be seen in the phase transition curve also shown in Figure \ref{fig:assortativity}.

As the density of nodes increases, assortativity decreases, making the network slightly disassortative. This behavior is related to the emergence of the giant component, formed by clusters, as shown in Figures S3 and S4 of the SM. In these clusters, nodes with many connections tend to link to nodes with few connections, reflecting the intrinsic heterogeneity of these networks. On the other hand, in the Brito et al. model, assortativity does not show this dependence on $\rho$. As the node density in this network increases, assortativity remains around $r \approx 0.4$, because the network is relatively homogeneous, with nodes having a similar number of connections. As a result, assortativity stays positive regardless of the density.

In this context, it can be questioned whether assortativity can indicate the phase transition in the Brito-Soares and Brito-Rozenfeld models. Although the point where $r=0$ is close to the critical density of the phase transition, they do not coincide exactly. 
However, assortativity can still serve as a qualitative indicator and guiding parameter of the phase transition, because as long as the giant component does not emerge, the network remains assortative. When the transition occurs, the network becomes slightly disassortative.

\begin{figure}[H]
    \centering
    \includegraphics[width=1\linewidth]{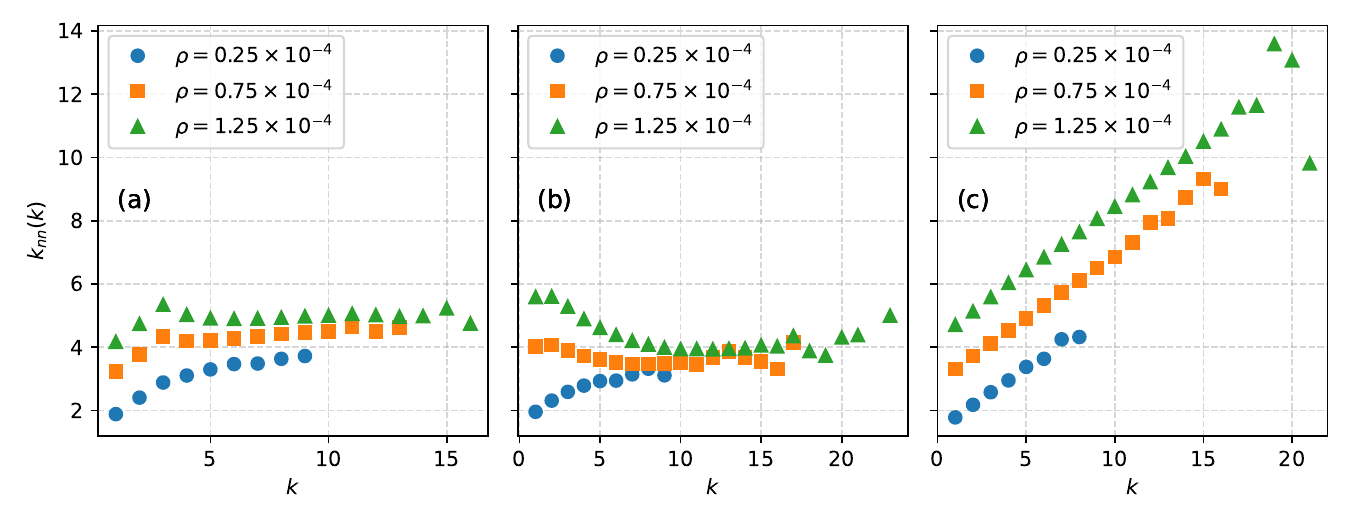}
    \caption{Average degree of nearest neighbors as a function of node degree for different $\rho$. In the Brito–Soares (a) and Brito–Rozenfeld (b) models, networks are assortative at low densities, uncorrelated at intermediate densities, and slightly disassortative at high densities, whereas in the Brito et al. (c) model, networks remain assortative.}
    \label{fig:knn}
\end{figure}

Moreover, Figure \ref{fig:knn} shows the average degree of nearest neighbors as a function of node degree for the three models at three different network densities. These results corroborate the findings reported in the previous analysis. For the Brito et al. model, the curves display a consistently positive slope, revealing an assortative structure. In contrast, for the Brito–Soares and Brito–Rozenfeld models, we observe the same behavior shown in Figures \ref{fig:assortativity}: at low densities, the networks are assortative; at intermediate densities, the assortativity remains nearly constant, indicating no correlation; and at higher densities, a slightly disassortative behavior begins to emerge.

\subsection{Properties of Real-World Fiber Optic Networks}

To compare the models with real data, we used six networks from the ITDK dataset, which represent inferred IP-layer connections. Although these do not directly correspond to physical fiber links -- they may correspond to LAN cables, Ethernet, or other physical media -- we map them using node geolocation and assume they could be realized as optical fibers. This is reasonable, as the distribution of future quantum network nodes would likely follow existing infrastructure. Since each network has a fixed number of nodes, we compare them with a fixed number of nodes $N$ for each proposed model. The results are summarized in Table \ref{tab:metricas_redes} that shows the number of nodes per continent in the analyzed real internet networks and provides a comparative analysis with the network models: Brito-Rozenfeld Network, Brito-Soares Network, and Brito et al. Network. Each of the six rows corresponds to a network topology for a different continent, characterized by its number of nodes ($N$) and density ($\rho$). For each of the four networks (real networks and proposed models), the table reports key structural metrics, including the average shortest path length ($\langle l \rangle$), the average clustering coefficient ($\langle C \rangle$), degree assortativity ($r$), and the density of links ($E/N$). This layout allows for a direct comparison of the real network's behavior against the theoretical models in each specific geographical scenario.

\begin{table}[h!]
\centering

\LARGE 
\renewcommand{\arraystretch}{2.0}

\caption{Topological metrics of photonic networks by continent. The networks contain 43,957 nodes in North America, 31,430 in Europe, 23,059 in South America, 21,317 in Asia, 5,540 in Oceania, and 2,032 in Africa.}
\label{tab:metricas_redes}
\resizebox{\textwidth}{!}{
\begin{tabular}{|c|c||c|c|c|c||c|c|c|c||c|c|c|c||c|c|c|c|}
\hline
\multirow{2}{*}{\textbf{Nodes (N)}} & \multirow{2}{*}{\textbf{Density ($\rho$)}} & \multicolumn{4}{c||}{\textbf{Real Network}} & \multicolumn{4}{c||}{\textbf{Brito-Rozenfeld Model}} & \multicolumn{4}{c||}{\textbf{Brito-Soares Model}} & \multicolumn{4}{c|}{\textbf{Brito et al. Model}} \\ \cline{3-18} 
 &  & $ \langle l \rangle $ & $ \langle C \rangle $ & $ R $ & $ E/N $ & $ \langle l \rangle $ & $ \langle C \rangle $ & $ R $ & $ E/N $ & $ \langle l \rangle $ & $ \langle C \rangle $ & $ R $ & $ E/N $ & $ \langle l \rangle $ & $ \langle C \rangle $ & $ r $ & $ E/N $ \\ \hline

43957 & $1.78 \times 10^{-3}$ & 5.1094 & 0.2733 & -0.065 & 2.94 & 15.3087 & 0.3951 & -0.1401 & 3.53 & 36.8978 & 0.4065 & 0.0067 & 1.95 & 13.2813 & 0.3971 & 0.4579 & 45.55 \\ \hline

31430 & $2.9848 \times 10^{-3}$ & 5.2909 & 0.2484 & -0.1101 & 3.98 & 16.2472 & 0.3840 & -0.1455 & 3.33 & 26.1107 & 0.4021 & 0.0285 & 1.97 & 8.5815 & 0.4009 & 0.5176 & 75.35 \\ \hline

23059 & $1.2925 \times 10^{-3}$ & 5.1597 & 0.4475 & 0.2957 & 10.34 & 20.9146 & 0.3660 & -0.2010 & 2.82 & 31.4944 & 0.4095 & -0.0171 & 1.93 & 11.8544 & 0.3985 & 0.4824 & 32.98 \\ \hline

21317 & $4.7817 \times 10^{-4}$ & 7.331 & 0.2984 & -0.0779 & 1.93 & 36.0163 & 0.3466 & -0.2438 & 2.23 & 47.3590 & 0.4091 & -0.0902 & 1.84 & 21.1506 & 0.3967 & 0.4233 & 8.38 \\ \hline

5540 & $6.4977 \times 10^{-4}$ & 4.6665 & 0.2840 & 0.2472 & 4.08 & 16.6429 & 0.3652 & -0.2373 & 2.49 & 22.4086 & 0.4047 & -0.0792 & 1.87 & 9.3468 & 0.4048 & 0.4446 & 16.36 \\ \hline

2032 & $6.6908 \times 10^{-5}$ & 4.3288 & 0.2964 & 0.1332 & 1.77 & 13.2351 & 0.2294 & -0.0852 & 1.09 & 8.1690 & 0.3687 & -0.0592 & 1.25 & 27.6299 & 0.3301 & 0.4374 & 1.71 \\ \hline
\end{tabular}%
}
\end{table}

Clearly, we do not expect the results for real networks to match those obtained from the models, even the heterogeneous ones. However, the main goal of this analysis is to observe the qualitative similarities between the models and the real networks, with respect to the photonic links. In this regard, we note that the link density of the real networks is very similar to that of the Brito-Rozenfeld and Brito-Soares models. On the other hand, the real networks exhibit slightly smaller average shortest path lengths compared to all models, possibly reflecting the small-world properties commonly found in technological networks \cite{network_barabasi}.  

Regarding the assortativity coefficient, the real networks show an slightly negative value, indicating a closer similarity to the Brito-Rozenfeld model than to the other two models. Finally, the average clustering coefficient is within the same range for all networks, indicating that all of them are slightly clustered.

\begin{figure}[h!]
        \centering
        \includegraphics[width=0.65\textwidth]{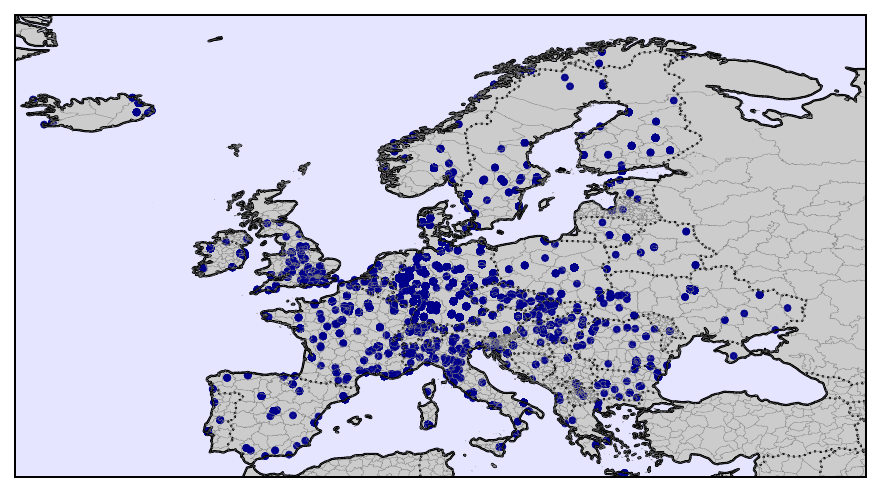}
        \caption{Geolocated node map of the real fiber optic network across Europe.}
    \label{fig:europe1}
\end{figure}

\begin{figure}[h!]
        \centering
    \begin{subfigure}[b]{0.32\textwidth}
        \centering
        \includegraphics[width=\textwidth]{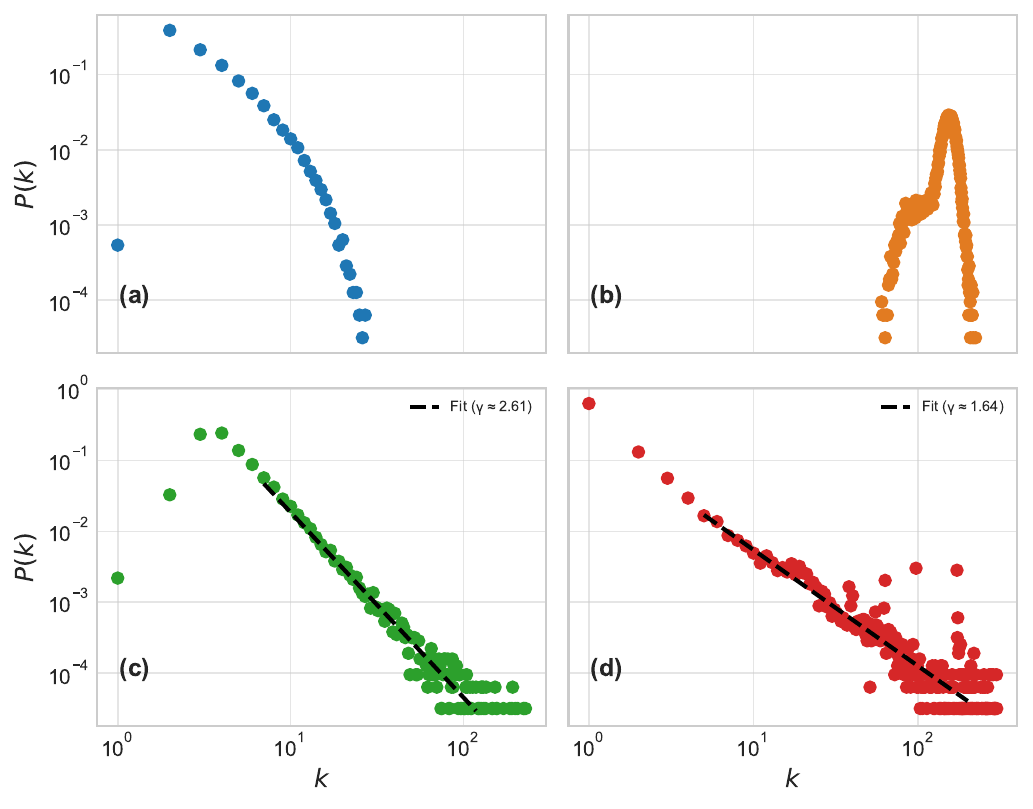}
        \captionsetup{labelformat=empty}
        \caption{Degree Distribution $P(k)$.}
    \end{subfigure}
    \hfill
    \begin{subfigure}[b]{0.32\textwidth}
        \centering
        \includegraphics[width=\textwidth]{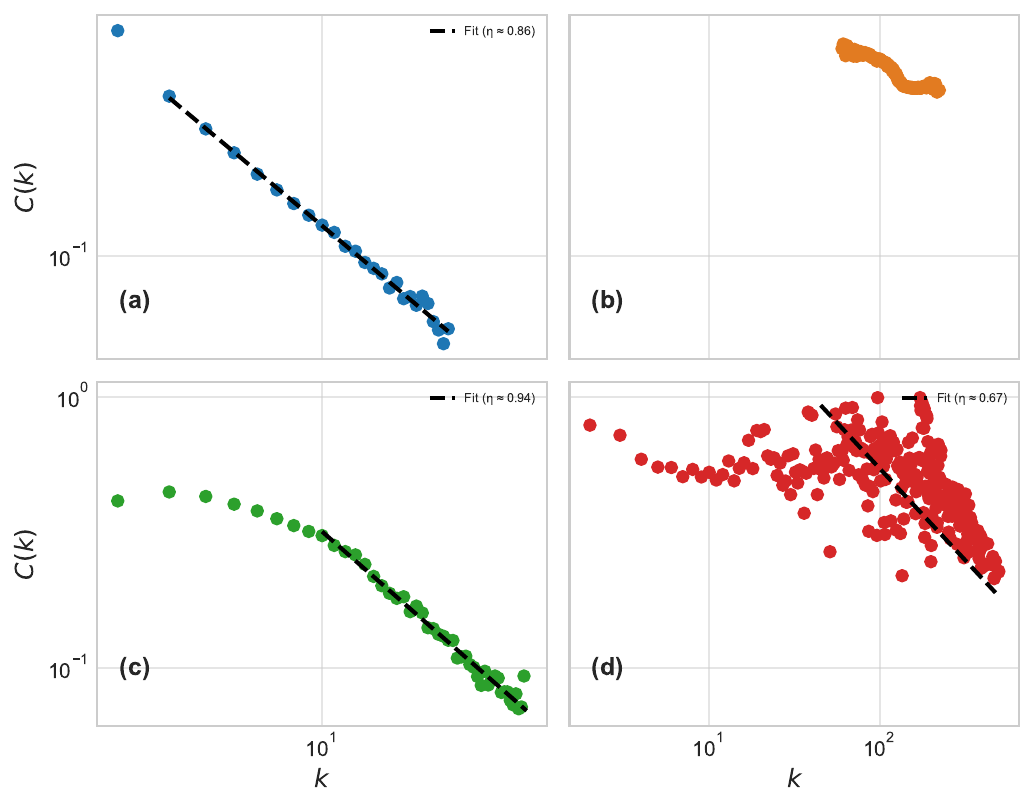}
        \captionsetup{labelformat=empty}
        \caption{Clustering Coefficient $C(k)$.}
    \end{subfigure}
    \hfill
    \begin{subfigure}[b]{0.32\textwidth}
        \centering
        \includegraphics[width=\textwidth]{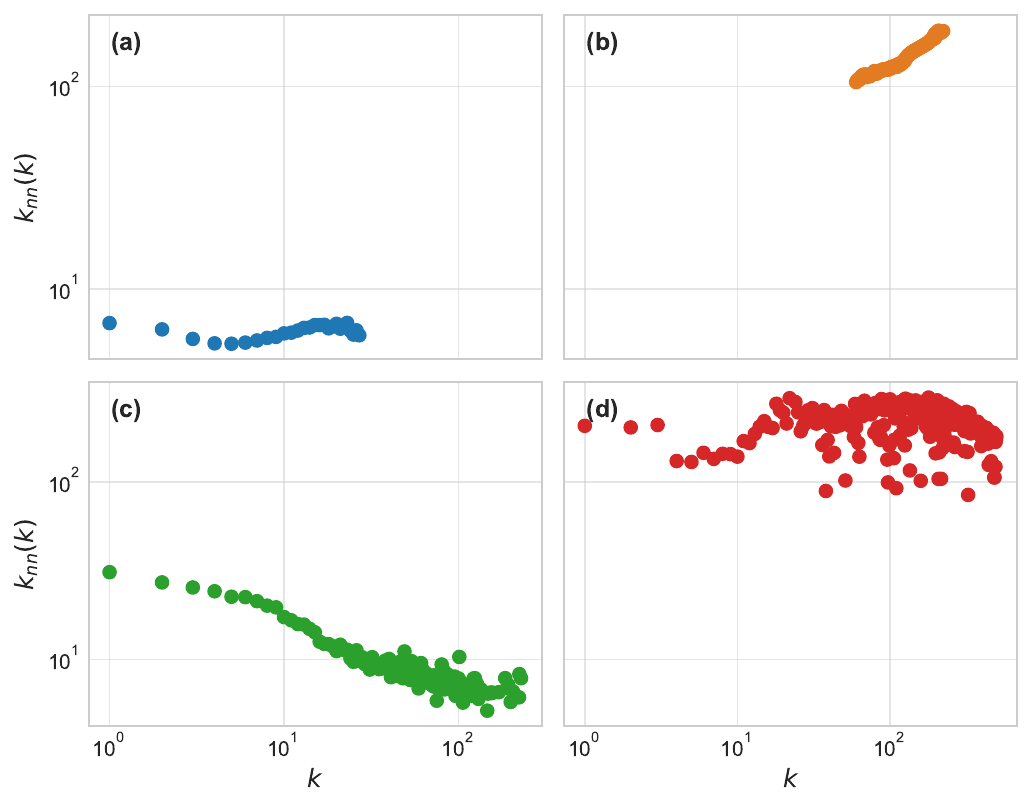}
        \captionsetup{labelformat=empty}
        \caption{K-Nearest Neighbors ($k_{nn}(k)$).}
    \end{subfigure}

    \caption{Analysis of the Europe network showing the degree distribution $P(k)$, clustering coefficient $C(k)$, and K-Nearest Neighbors ($k_{nn}(k)$) for the different networks. In each subfigure we show the comparison of  (a) Brito-Soares , (b) Brito et al., (c) Brito-Rozenfeld, (d) Real Network.}
    \label{fig:europe2}
\end{figure}

In Figure \ref{fig:europe1}, we show a map of Europe with the geolocated nodes, representing the vertex locations of the underlying network on which the photonic network could be deployed. Figure \ref{fig:europe2} presents a comparison of this real network with the three models under investigation. We can observe that the degree distribution of the real network is clearly more similar to the Brito-Rozenfeld model, as expected since most real networks are heterogeneous \cite{network_barabasi}.

The clustering coefficient as a function of degree $k$ also exhibits a hierarchical behavior, following a power-law decay, which is consistent with both the Rozenfeld-Brito and Soares-Brito models and reflects a property commonly found in real internet networks \cite{barabasi_hierarchical,internet_power_law}. Finally, the average degree of nearest neighbors shows behavior between no correlation and slightly disassortative, also similar to the heterogeneous models.  

Overall, these analyses strongly support that our newly proposed heterogeneous models describe fiber optic networks more accurately than the homogeneous models previously reported in the literature \cite{brito2020statistical,brito2021satellite,oliveira2024statistical}. The plots for the other five real networks can be found in Section S4.1 of the SM. The analysis for these networks is largely the same as that performed for the European network.


\section{Conclusions}\label{sec:conclusions}

In this work, new models were proposed for the structure of the quantum internet, taking into account the heterogeneity present in fiber optic networks. Unlike existing models in the literature \cite{brito2020statistical,oliveira2024statistical}, which assume a homogeneous distribution of connections, here we analyzed the impact of heterogeneity in the distribution of connections on the statistical properties of the quantum network. This approach allows for a more realistic description of the quantum communication infrastructure.

The results show similarities with previous studies, but also reveal significant differences. Although the Brito et al. model \cite{brito2020statistical} exhibits shorter average path lengths due to its higher link density (when comparing the same number of nodes and density in the circular space), the heterogeneous Brito-Soares and Brito-Rozenfeld models achieve comparable efficiency with fewer connections, as high-degree nodes act as shortcuts linking distant regions of the network. This highlights the advantage of heterogeneous structures in optimizing connectivity while minimizing the number of links required. 
This result is relevant because the average shortest path length is directly related to the efficiency of the entanglement distribution in the  network \cite{oliveira2024statistical}. In addition, the assortativity of these networks was analyzed, highlighting that, for the new models, this property can serve as an indicator of the phase transition to a highly connected state, in which the network becomes slightly disassortative. 
Furthermore, we also find that for the Brito et al. model, $C(k)$ remains roughly constant, while the Brito-Soares and Brito-Rozenfeld models show a power-law decay, as observed in the internet, indicating that the proposed models better capture hierarchical network features.

To conclude, we analyzed the distribution of connections in infrastructure networks across six continents, consisting of large and robust networks. Although the results for real networks do not exactly match those of the models, the analysis reveals strong qualitative similarities. In particular, the degree distribution and link density of real networks closely resemble those of the Brito-Rozenfeld and Brito-Soares models. The real networks exhibit slightly shorter average path lengths, reflecting small-world properties, while their assortativity presents a behavior between no correlation and slightly disassortative, similar to the Brito-Rozenfeld model. Finally, the average clustering coefficient is comparable across all networks, indicating slight clustering; moreover, the clustering as a function of degree follows a power-law decay, consistent with the heterogeneous models and hierarchical models representing internet structures in previous studies \cite{internet_power_law}. This analysis highlights that the proposed heterogeneous models best capture the structure of the emerging quantum internet, providing guidance for developing more efficient architectures for global quantum communication.
~\\

\section*{Competing Interests}

The authors declare that they have no competing interests.
~\\

\section*{Acknowledgment}

The authors would like to thank Dr. Guilherme Henrique da Silva Costa and Dr. Sergio Martins de Souza for their careful review and valuable contributions. OJS and NBS acknowledge the financial support from CNPq 
(Conselho Nacional de Desenvolvimento Científico e Tecnológico) and  ASM acknowledges the financial support from CNPq/Fapemig APQ-06591-24 (Conselho Nacional de Desenvolvimento Científico e Tecnológico and Fundção de Amparo à Pesquisa do Estado de Minas Gerais). 

\section*{Data availability}

The codes used in this work are available on github at the link: 
\\
\url{https://github.com/otavio-silveira/qnetwork-models}.
~\\

\section*{Author Contributions}

AM developed the original ideas; All authors designed the study; AM and SS supervised the development of the experiments; OS wrote the formalisms; OS and NS collected, curated, and integrated the raw data; OS and NS performed the analysis; All authors analysed the results and wrote the manuscript; OS and NS prepared the graphics; all authors read, revised  and approved the final version of the manuscript. 
~\\

\clearpage
\appendix
\renewcommand{\thesection}{S\arabic{section}} 
\section*{Supplementary Material}
\addcontentsline{toc}{section}{Supplementary Material}

\setcounter{section}{0} 

\section{Characterization of Complex Networks}
\label{sec:supp_characterization}

A network consists of a set of $N$ nodes, representing the components of the system under study, which interact with each other through links. Each node has a specific number of links, called a degree, represented by $k_i$ for a given node $i$. The degree distribution of a network is the function $P(k)$ ( properly normalized) that describes the probability of a node having exactly $k$ connections \cite{network_barabasi}.

As the density of nodes in the network increases, so does the probability of connections being formed between them. Initially, the network is made up of several isolated sub-networks, where the nodes interact only within small disconnected groups. However, as the network becomes denser, the availability of nearby nodes increases, facilitating the sharing of entangled photons via quantum channels and enabling connections that were previously impractical due to large distances. This process leads to a phase transition, in which the network goes from a fragmented state to one in which a significant fraction of the nodes are connected, giving rise to a giant component - the largest connected sub-network. The highly connected state can be characterized by the fraction $N_G/N$, between the number of nodes $N_G$ in the giant component and the total number of nodes $N$ in the network. Initially, this fraction is small because the network is fragmented, but it grows rapidly when the giant component emerges. 

There are some basic properties of complex networks that are very used to characterize them, such as:

\begin{itemize}

\item[] {\bf Average shortest path length:}
~\\

The average shortest path length of a network, denoted by $\left \langle l \right \rangle$, measures the average distance between all pairs of nodes. For a network with $N$ nodes, $d_{ij}$ is defined as the shortest distance, in number of links, between nodes $i$ and $j$. Thus, the average shortest  path length is given by:

\begin{equation*}
    \left \langle l \right \rangle = \frac{1}{N(N-1)}\sum_{i \neq j} d_{ij},
\end{equation*}

where the sum goes through all distinct pairs of nodes \cite{network_barabasi}. In the networks presented here, $\left \langle l \right \rangle$ is calculated only for the giant component, because in disconnected networks, the distance between nodes can be $d_{ij} = \infty$. A network is said to have the small-world property when $\left \langle l \right \rangle$ scales logarithmically with $N$, meaning that most nodes can be reached from any other in just a few links.
~\\

\item[] {\bf Average Clustering Coefficient:}
~\\

The average clustering coefficient $\left \langle C \right \rangle$ measures the degree of clustering between nodes in the network. For a node $i$, its clustering coefficient $C_i$ is the fraction of pairs of neighbors that are also connected to each other, given by

\begin{equation*}
    C_i=\frac{2L_i}{k_i(k_i-1)},
\end{equation*}

where $L_i$ is the number of triangles formed with node $i$ and $k_i$ its degree. The average clustering coefficient is obtained by taking the average of $C_i$ over all the nodes. High values of $\left \langle C \right \rangle$ indicate highly locally connected networks, while low values suggest more sparse structures \cite{Mata2020}.
~\\

\item[] {\bf Clustering Coefficient as a Function of Degree:}
~\\

This is another way to quantify the local clustering in a network, now considering the connectivity degree of the nodes. The function $C(k)$ is defined as the average clustering coefficient of all nodes with degree $k$:

\begin{equation*}
    C_k = \frac{1}{N_k}\sum_{i \in \{v: k_v = k\}} C_i
\end{equation*}

where $N_k$ is the number of nodes with degree $k$. This measure reveals structural patterns such as hierarchy or modularity in the network \cite{network_barabasi}.
~\\

\item[] {\bf Assortativity:}
~\\

The assortativity of a network quantifies the tendency for nodes with similar degrees to be connected with each other. It is measured by the degree assortativity coefficient $r$, defined as the Pearson correlation between the degrees of connected nodes \cite{assort_newman}. This coefficient varies between $r=1$, when nodes of similar degree tend to connect (assortative network), and $r=-1$, when nodes of high degree prefer to connect to nodes of low degree (disassortative network). If $r=0$ means that there is no correlation degree in the network \cite{network_barabasi}.
~\\

\item[] {\bf Average Degree of Nearest Neighbors as a Function of Node Degree:}
~\\

It provides a complementary measure to the Pearson correlation coefficient for quantifying network assortativity.
The function $k_{nn}(k)$ expresses the average degree of the neighbors of nodes that have a given degree $k$, capturing how the connectivity of a node relates to the connectivity of its neighbors. It is defined as:

\begin{equation*}
    k_{nn}(k) = \frac{1}{N_k}\sum_{i \in \left\{ v:k_v=k \right\}} \left( \frac{1}{k}\sum_{j \in \mathcal{N}_i} k_j \right),
\end{equation*}

where $N_k$ is the number of nodes with degree $k$, $\mathcal{N}_i$ denotes the set of neighbors of node $i$, and $k_j$ is the degree of neighbor $j$. For each node with degree $k$, one computes the average degree of its neighbors, and then takes the mean over all such nodes.

The behavior of $k_{nn}(k)$ provides insights into the degree correlations of the network. When $k_{nn}(k)$ increases with $k$, the network exhibits assortative properties, where high-degree nodes tend to connect to other high-degree nodes. In contrast, when $k_{nn}(k)$ decreases with $k$, the network is disassortative, meaning that high-degree nodes tend to be connected to low-degree ones. A flat profile of $k_{nn}(k)$, with no dependence on $k$, indicates the absence of degree correlations \cite{network_barabasi,assort_newman}. This function offers a more detailed characterization of connectivity patterns than the global assortativity coefficient and is especially useful for identifying hierarchical or modular structures in complex networks.

\end{itemize}

\section{Models for Quantum Networks}\label{subsec:models}
\label{sec:supp_models}

\begin{figure}[htb]
    \centering
    \includegraphics[width=0.8\linewidth]{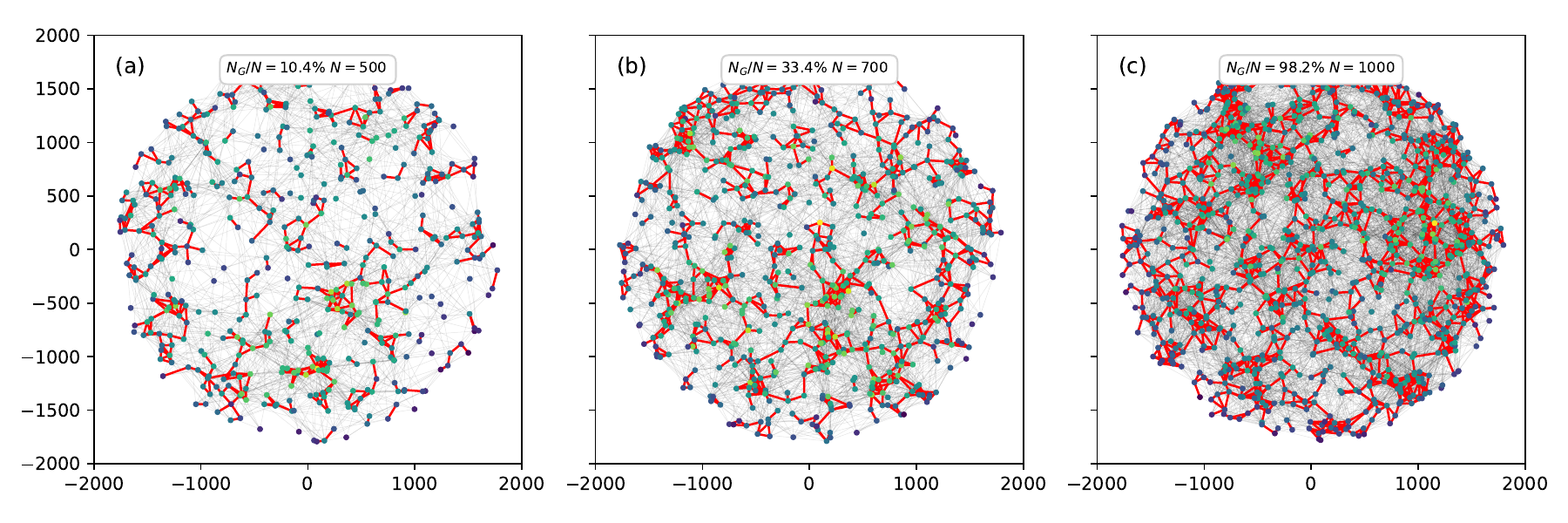}
    \caption{Samples of the model proposed by Brito et al \cite{brito2020statistical}. The optical fibers are represented by the gray edges. Blue nodes have few connections, while yellow nodes have many. The red connections represents the quantum network, indicate nodes that share pairs of entangled photons. The ratio $N_G/N$, between the number of nodes in the largest component of the network and the total number of nodes, shows the network transitions, from left to right, from a weakly connected network to a highly connected one as the density of nodes increases.}
    \label{fig:samurai_samples}
\end{figure}

\begin{figure}[htb]
    \centering
    \includegraphics[width=0.8\linewidth]{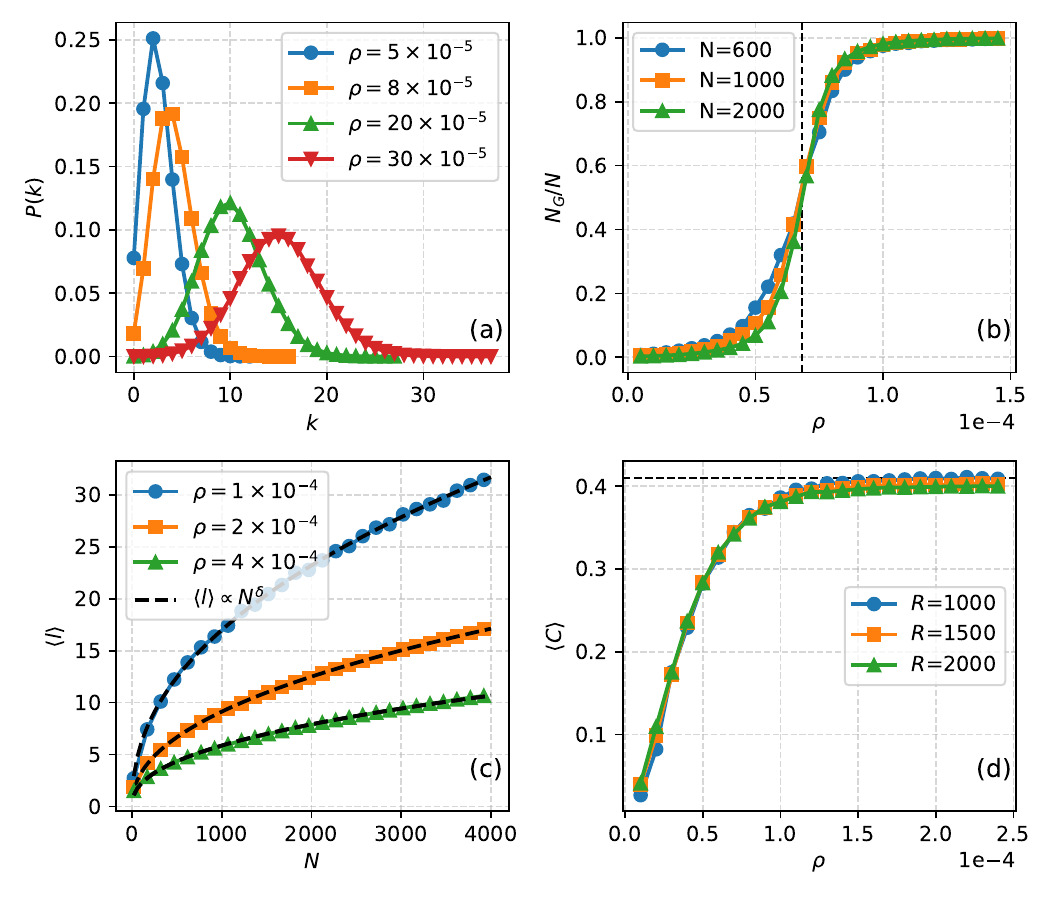}
    \caption{Main properties of photonic networks running on top of the fiber-optic network model proposed by Brito et. al. (a) Its degree connectivity distribution follows a Poisson distribution, where it can be observed that, as the density of nodes increases, the network becomes increasingly homogeneous. (b) The network presents a phase transition at a critical density close to $6.82 \times 10^{-5}$. (c) The network does not present a small-world property, with $\left\langle l \right\rangle \propto N^{\delta}$, where $\delta = 0.45$. (d) The network is vary aggregated, with $\left \langle C \right \rangle \approx 0.41$.}
    \label{fig:samurai_properties}
\end{figure}

Figure \ref{fig:samurai_samples} shows samples of the photonic networks operating on the fiber-optic network model proposed by Brito et al. \cite{brito2020statistical}. A central feature of this model is that its connectivity distribution between fibers follows a Poisson distribution, as shown in Figure \ref{fig:samurai_properties}(a), similar to random graphs. This implies that most nodes have a similar number of connections, resulting in a more homogeneous network. In addition, the model exhibits a second-order phase transition as a function of node density, moving from a loosely connected network to a highly connected one (see \ref{fig:samurai_properties}(b)).

However, unlike random graphs, the model shows a different behavior in the average shortest path length $\left\langle l \right\rangle$ between nodes. While networks with the small-world property follow $\left\langle l \right\rangle \propto \ln N$, in this case, the average shortest path length grows faster, following $\left\langle l \right\rangle \propto N^{\delta}$, where $\delta \approx 0.45$, as shown in Figure \ref{fig:samurai_properties}(c). Furthermore, as shown in Figure \ref{fig:samurai_properties}(d), the clustering coefficient of this model is $\left \langle C \right \rangle \approx 0.41$, indicating that these networks are very aggregated.

\begin{figure}[htb]
    \centering
    \includegraphics[width=0.8\linewidth]{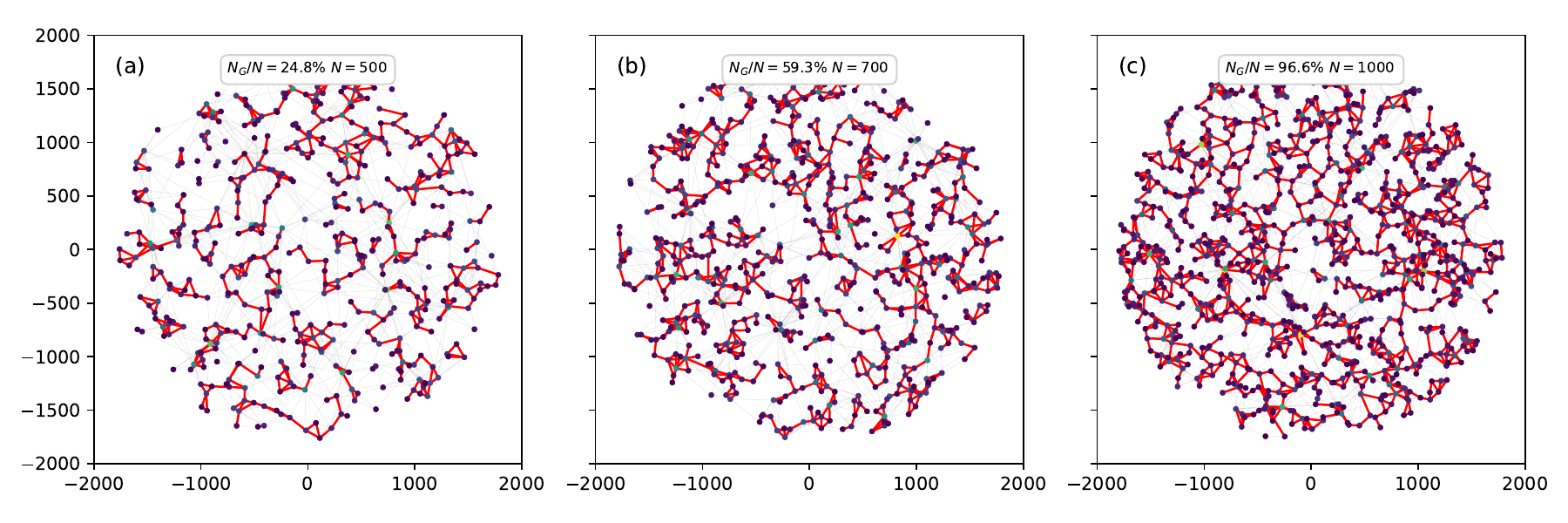}
    \caption{Samples of the preferential attachment model based on Soares et al. \cite{soares2005preferential}, where the phase transition is emphasized, for $\alpha_A = 5$ and $m=3$. The red connections represents the photonic network. The heterogeneity is evident in the presence of many nodes with few connections and few nodes with many connections.}
    \label{fig:soares_samples}
\end{figure}

Figure \ref{fig:soares_samples} presents examples of photonic networks running on top of the Brito-Soares fiber-optic network model  \cite{soares2005preferential} for $\alpha_A = 5$, highlighting its phase transition. The color pattern of the nodes shows the heterogeneity of the network: most of the nodes are in blue (few connections), with a few in green and even fewer in yellow (many connections).

\begin{figure}[htb]
    \centering
    \includegraphics[width=0.9\linewidth]{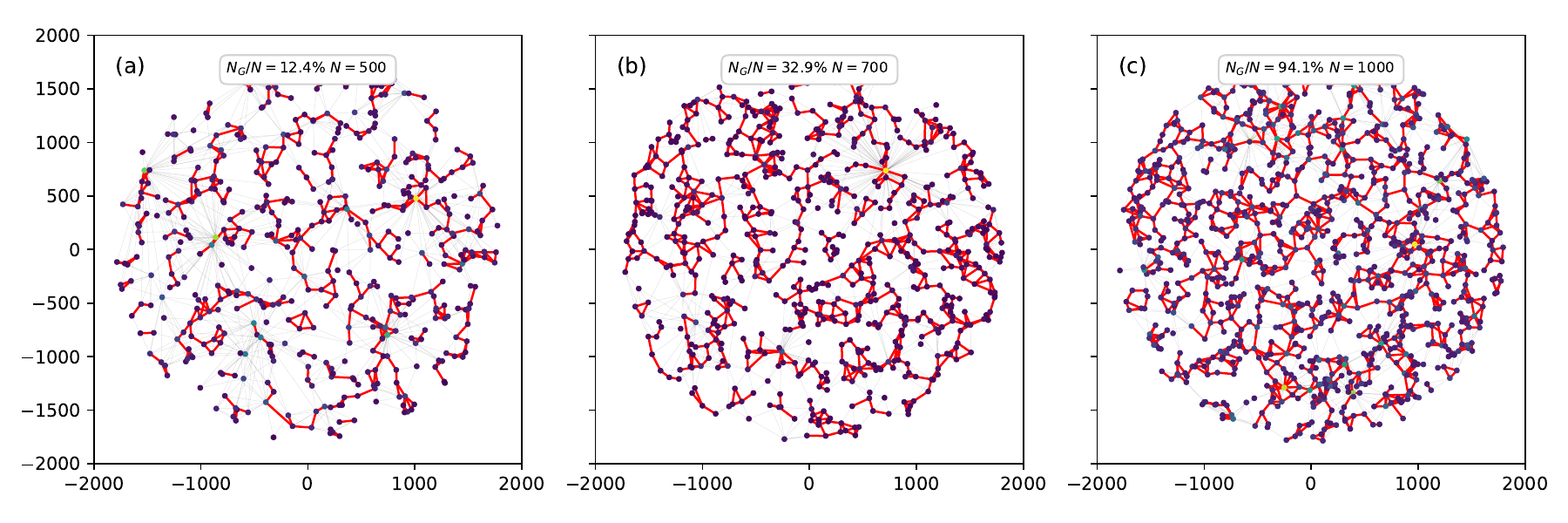}
    \caption{Samples of the scale-free model based on Rozenfeld et. al. \cite{rozenfeld2002scale} highlighting the phase transition for $m=3$, $K=1 \times 10^6$, $\lambda=3$ and $A=100$. The red connections represents the quantum network. The effect of the scale-free degree distribution is to form a highly heterogeneous network.}
    \label{fig:scale_free_samples}
\end{figure}

Figure \ref{fig:scale_free_samples} illustrates photonic networks constructed on the Brito-Rozenfeld network model \cite{rozenfeld2002scale} for $\lambda = 3$, showing its phase transition. We can see a similar pattern as in the growth and preferential attachment model: most nodes have few connections, while a few nodes contain many connections. Also important is the impact of the connection radius present in Eq (3) of the main text, $r(k)=A\sqrt{k}$: even though the parameter $A$ is limited, nodes with high values of $k$ are able to establish long-distance connections in the network.

\section{Data Acquisition and Preprocessing for Real Fiber Optic Networks}
\label{sec:supp_preprocessing}
In order to evaluate the applicability of the proposed network models, we utilized real-world internet network data for comparative purposes.

1. The dataset used for the analysis was the \textbf{August 2020 Internet Topology Data Kit (ITDK)}\footnote{\url{https://www.caida.org/catalog/datasets/internet-topology-data-kit/release-2020-08/}}, maintained by the Center for Applied Internet Data Analysis (CAIDA). The dataset contains global internet (fiber optic) topology data. Specifically, two types of data files were used:
\begin{itemize}
    \item \textbf{Geolocated Nodes:} A text file containing geographic and metadata information for each node (router) identified in the measurement. Each line includes the node ID, its location (continent, country, city), geographic coordinates (latitude and longitude), and the source of the geolocation information.
    \item \textbf{Links between Nodes:} A text file describing the inferred connections between nodes. The format of this file is notably complex, as each line can represent a ``hyperedge,'' i.e., a link that connects multiple nodes in a clique topology.
\end{itemize}

2. For the extraction and filtering of the raw data, custom parsing routines were developed. A function using Python's Regular Expressions library (\textbf{re})\footnote{\url{https://docs.python.org/3/library/re.html}} was utilized for the extraction of geolocated nodes. This approach allowed for the precise extraction of each data field from every line. For the links file, it was necessary to construct a function to generate all unique pairs of connections, correctly translating the topology into an explicit edge list.

3. Once the raw data was read and transformed into two dataframes — one for geolocated nodes and one for links — a cross-validation step was performed to associate the links with geolocated nodes. The objective was to create a consistent data set in which each node had a known geographic location and each link connected two valid nodes. The process involved three steps:

\begin{itemize}
    \item First, a list was created containing all unique node IDs that participated in at least one connection in the links file (the set of ``active nodes'').
    \item Subsequently, the geolocated nodes DataFrame was filtered to retain only the entries whose node ID was present in the set of active nodes. This ensures that the final list of nodes contains only routers that are both geolocated and part of the connectivity topology.
    \item Finally, the links Data Frame were filtered a second time, keeping only the edges where both nodes (source and destination) were present in the final list of filtered geographic nodes.
\end{itemize}

4. The final stage consisted of enriching the link list with geographic data and constructing the final Data Frames to be used for analysis. 

\begin{itemize}
    \item The filtered links Data Frame were combined with the filtered nodes Data Frame through a merge operation. This join was performed twice: once to attach the data of the source node and a second time to attach the data of the destination node. The result was a single primary Data Frame where each row represented a complete link, containing the pair of connected nodes and their respective geographic information (latitude, longitude, continent, country, and city).
    \item Finally, from this primary Data Frame, individual Data Frames were segmented and exported for each continent, containing only the intracontinental links and their associated geographic data.

\end{itemize}

\section{Additional Results}

\begin{figure}[htb]
    \centering
    \includegraphics[width=0.85\linewidth]{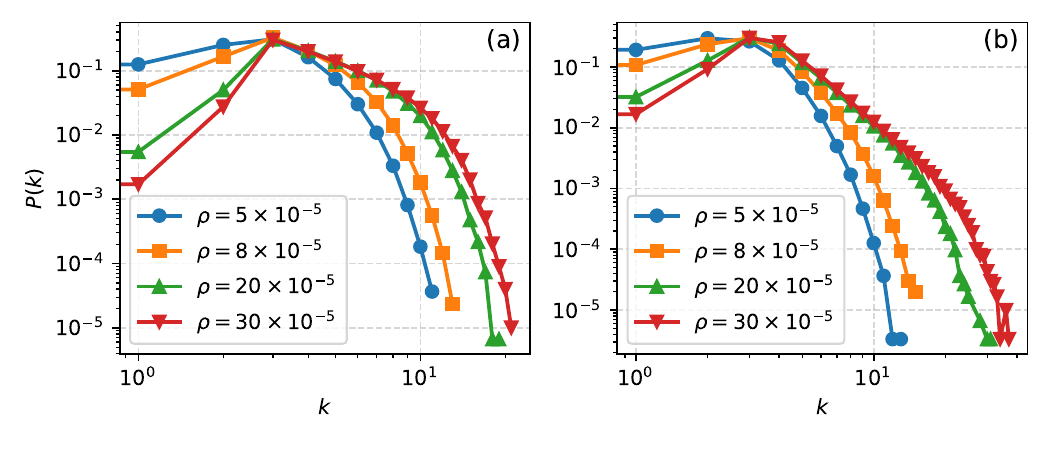}
    \caption{Degree distribution for different densities, with $N = 3000$. (a) Brito-Soares model and (b) Brito-Rozenfeld model. The distribution varies with density in both models.} 
    \label{fig:degree_distrib_for_rho}
\end{figure}

\begin{figure}[htb]
    \centering
    \includegraphics[width=0.85\linewidth]{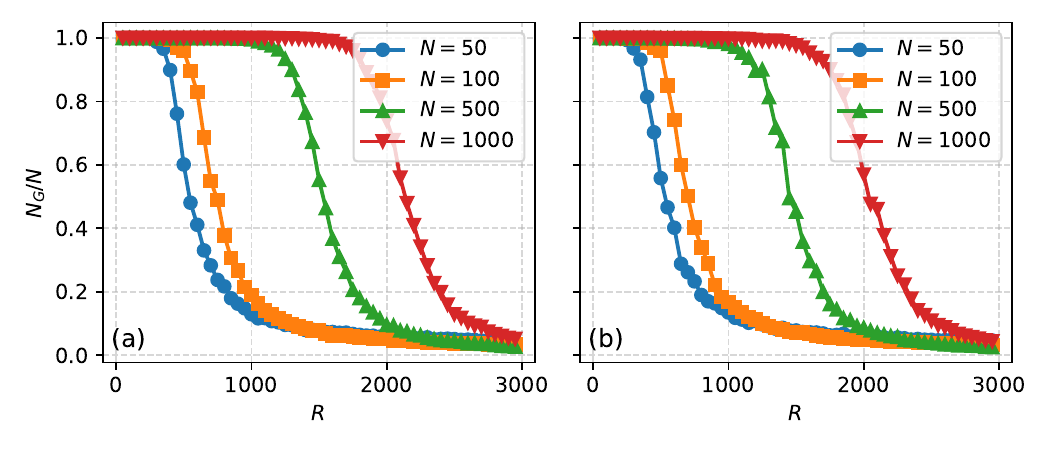}
    \caption{Phase transition as a function of $R$ for different numbers of nodes. (a) Brito-Soares model and (b) Brito-Rozenfeld model.}
    \label{fig:phase_transition_for_R}
\end{figure}

Figure \ref{fig:degree_distrib_for_rho} shows the degree distribution for different node densities for both Brito-Soares and Brito-Rozenfeld model. The general behavior is similar to that described in the main text about the degree distribution in Figure 2. However, as found in the work of Brito et al.\cite{brito2020statistical}, there is a dependence of the distribution on density. This can be explained because by keeping the number N of nodes fixed and increasing the density, we are, in fact, reducing the radius R of the disk in which the nodes are allocated. Therefore, they become closer, and by the rule of preferential attachment, there is a greater chance of having more connections (higher values of $k$ in the histogram appear more frequently). Figure \ref{fig:phase_transition_for_R} shows the phase transition, for both heterogeneous model, for different values of $R$, making it possible to estimate the minimum number of nodes needed to form a highly connected network in a specific area with an arbitrary $R$.

\subsection{Real Internet Networks}

In this section, we present the results for the photonic networks running on top of the real infrastructure network of five additional continents. The analysis is largely the same as that discussed in the main text, confirming that the Ronzefeld-Brito and Soares-Brito heterogeneous models are more suitable for describing real networks compared to previously proposed homogeneous models in the literature.

\begin{figure}[h!]
    \centering
    \begin{subfigure}[c]{0.6\textwidth}
        \centering
        \includegraphics[width=\textwidth]{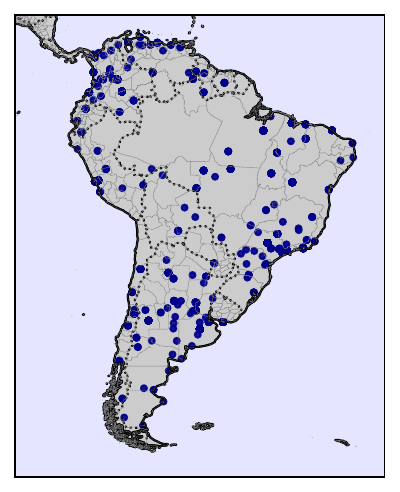}
        \captionsetup{labelformat=empty}
        \caption{Map of geolocated nodes.}
    \end{subfigure}
    \hfill
    \begin{subfigure}[c]{0.35\textwidth}
        \centering
        \includegraphics[width=\textwidth]{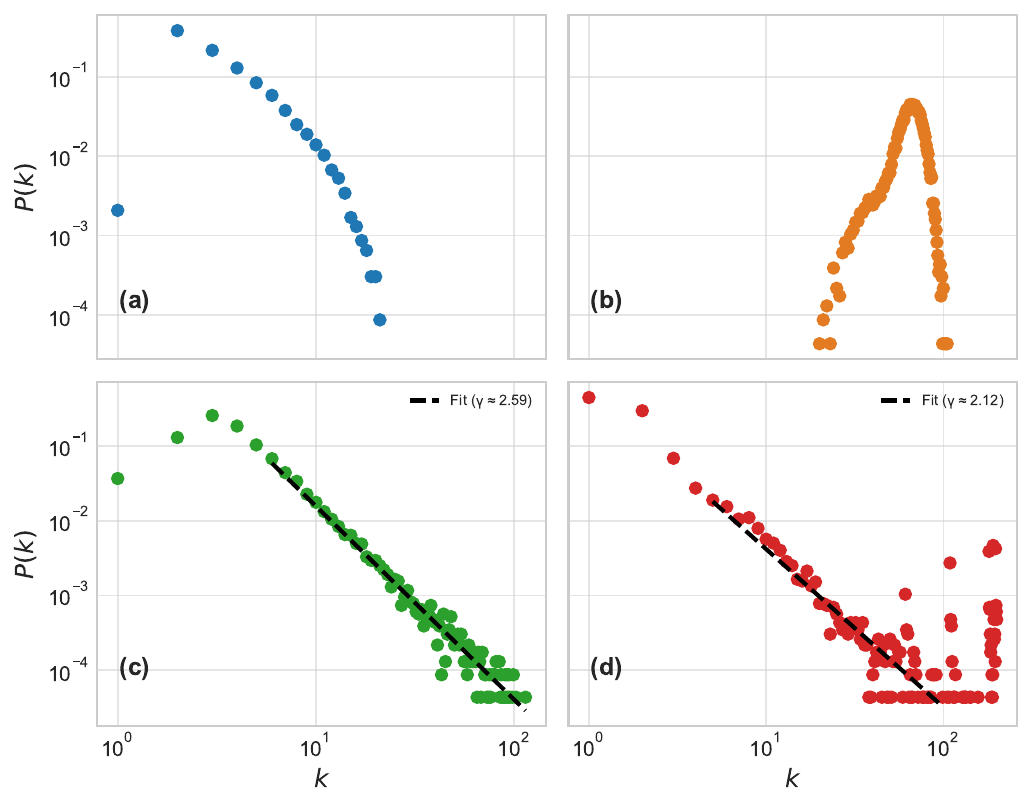}
        \captionsetup{labelformat=empty}
        \caption{Degree Distribution $P(k)$.}
        
        \vskip 0.3cm 
        \includegraphics[width=\textwidth]{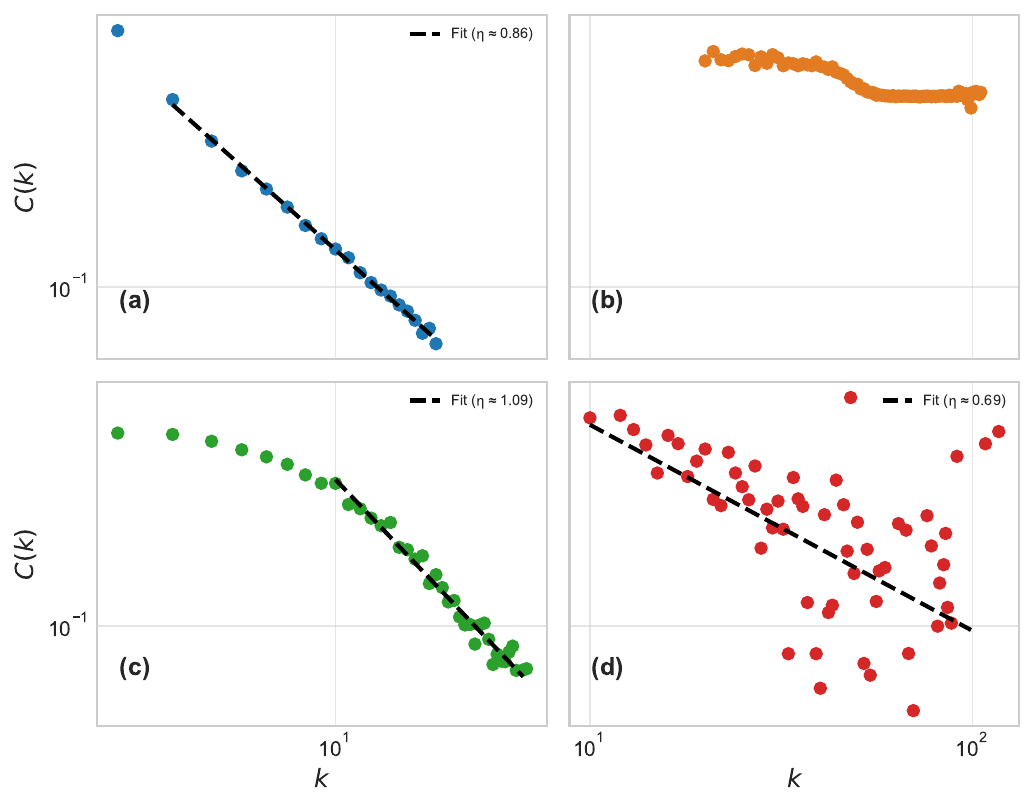}
        \captionsetup{labelformat=empty}
        \caption{Clustering Coefficient $C(k)$.}
        
        \vskip 0.3cm
        \includegraphics[width=\textwidth]{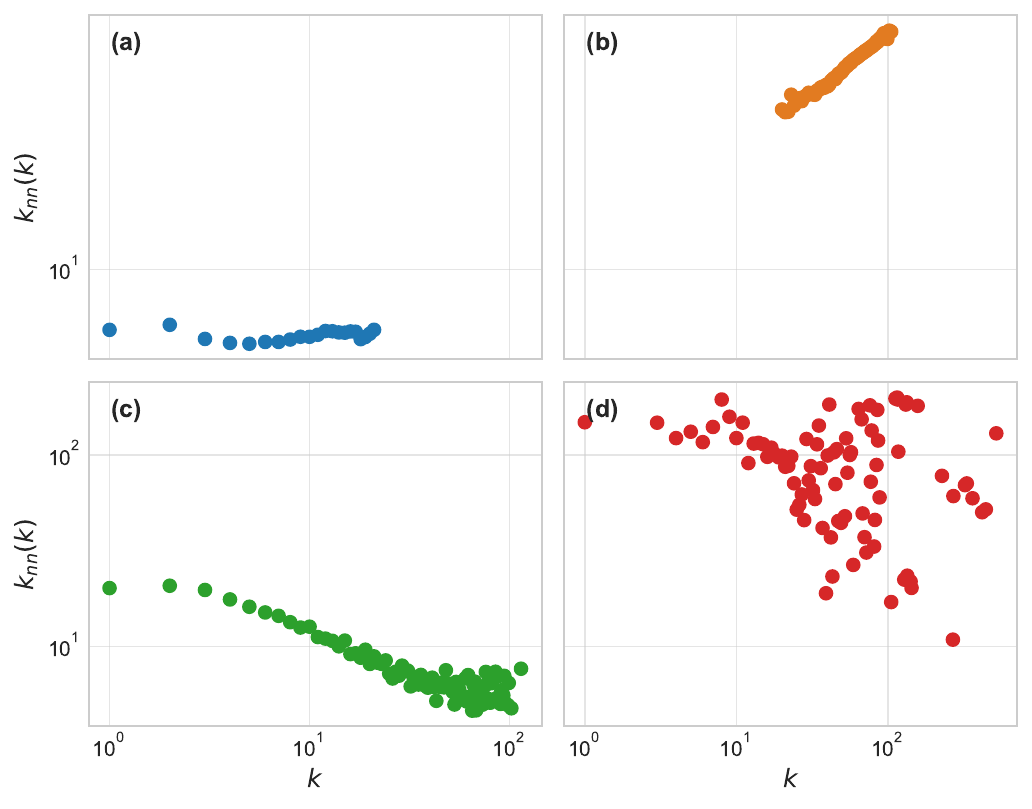}
        \captionsetup{labelformat=empty}
        \caption{K-Nearest Neighbors ($k_{nn}(k)$).}
    \end{subfigure}

    \caption{Analysis of the South America network. The left panel shows the map of geolocated nodes, and the right panels show the degree distribution $P(k)$, clustering coefficient $C(k)$, and K-Nearest Neighbors ($k_{nn}(k)$) for the different networks. In each subfigure we show the comparison of  (a) Brito-Soares , (b) Brito et al., (c) Brito-Rozenfeld, (d) Real Network.}
    \label{fig:southamerica}
\end{figure}

\begin{figure}[h!]
    \centering
    \begin{subfigure}[c]{0.6\textwidth}
        \centering
        \includegraphics[width=\textwidth]{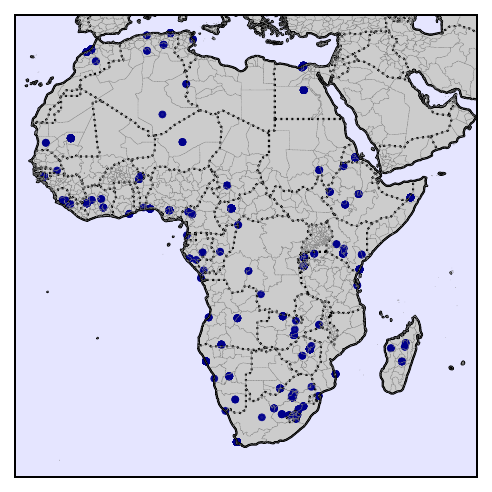}
        \captionsetup{labelformat=empty}
        \caption{Map of geolocated nodes.}
    \end{subfigure}
    \hfill
    \begin{subfigure}[c]{0.35\textwidth}
        \centering
        \includegraphics[width=\textwidth]{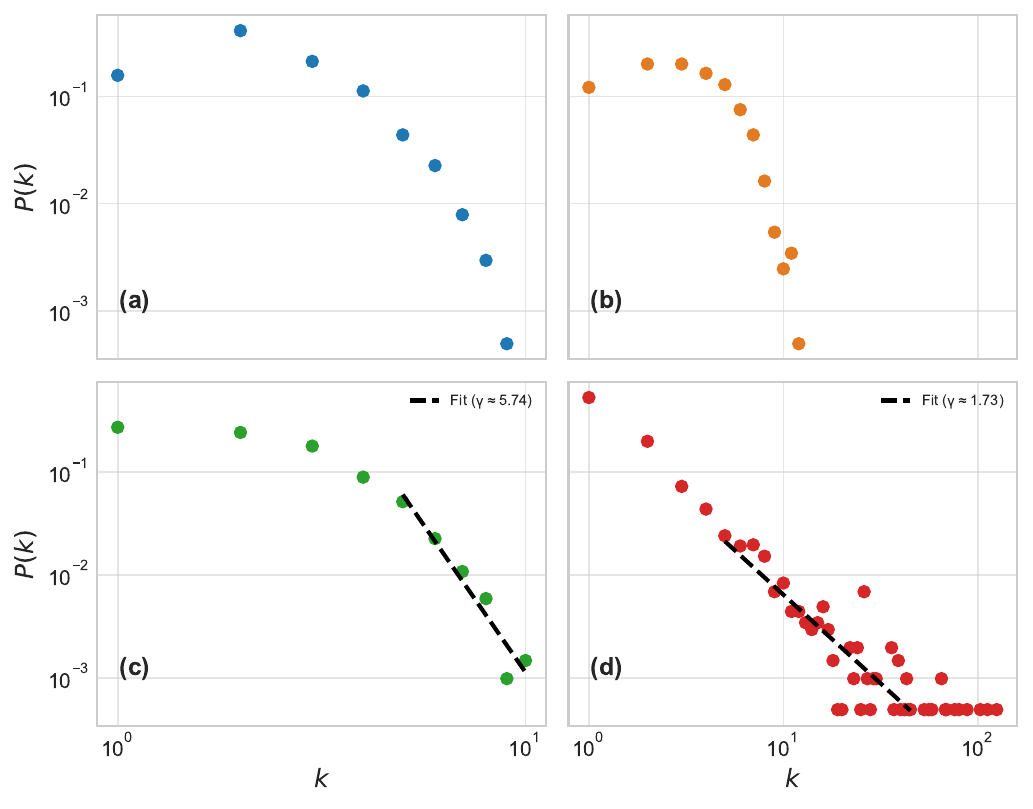}
        \captionsetup{labelformat=empty}
        \caption{Degree Distribution $P(k)$.}
        
        \vskip 0.3cm 
        \includegraphics[width=\textwidth]{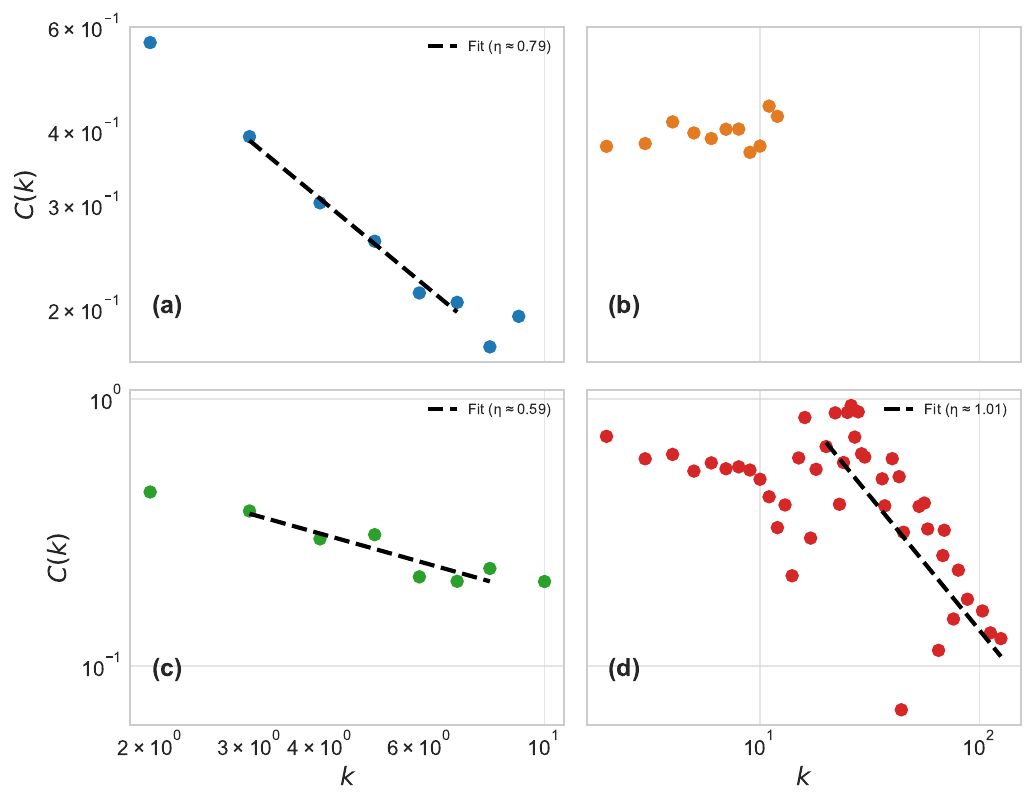}
        \captionsetup{labelformat=empty}
        \caption{Clustering Coefficient $C(k)$.}
        
        \vskip 0.3cm
        \includegraphics[width=\textwidth]{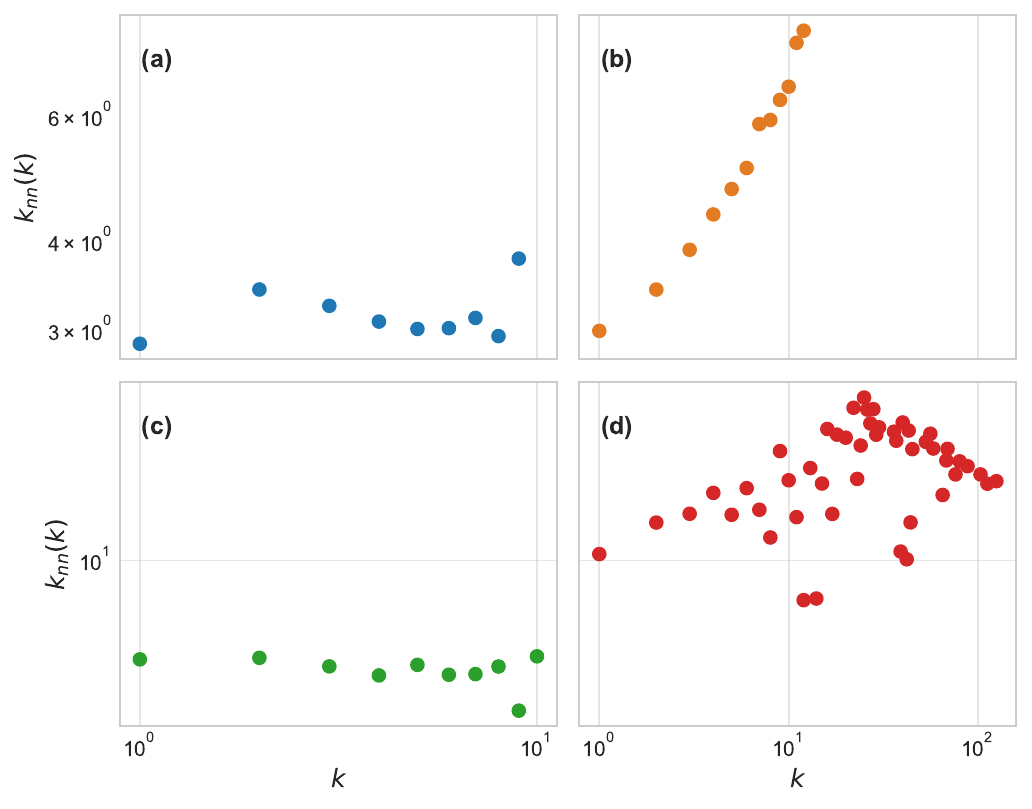}
        \captionsetup{labelformat=empty}
        \caption{K-Nearest Neighbors ($k_{nn}(k)$).}
    \end{subfigure}

    \caption{Analysis of the Africa network. The left panel shows the map of geolocated nodes, and the right panels show the degree distribution $P(k)$, clustering coefficient $C(k)$, and K-Nearest Neighbors ($k_{nn}(k)$) for the different networks. In each subfigure we show the comparison of  (a) Brito-Soares , (b) Brito et al., (c) Brito-Rozenfeld, (d) Real Network.}
    \label{fig:africa}
\end{figure}

\begin{figure}[h!]
    \centering
    \begin{subfigure}[b]{0.9\textwidth}
        \centering
        \includegraphics[width=\textwidth]{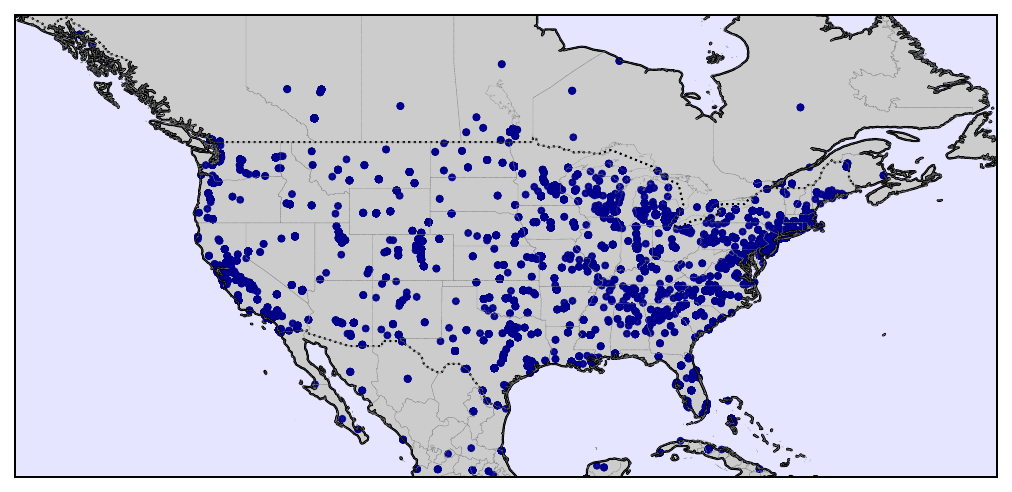}
        \captionsetup{labelformat=empty}
        \caption{Map of geolocated nodes.}
    \end{subfigure}

    \vskip 0.3cm

    \begin{subfigure}[b]{0.32\textwidth}
        \centering
        \includegraphics[width=\textwidth]{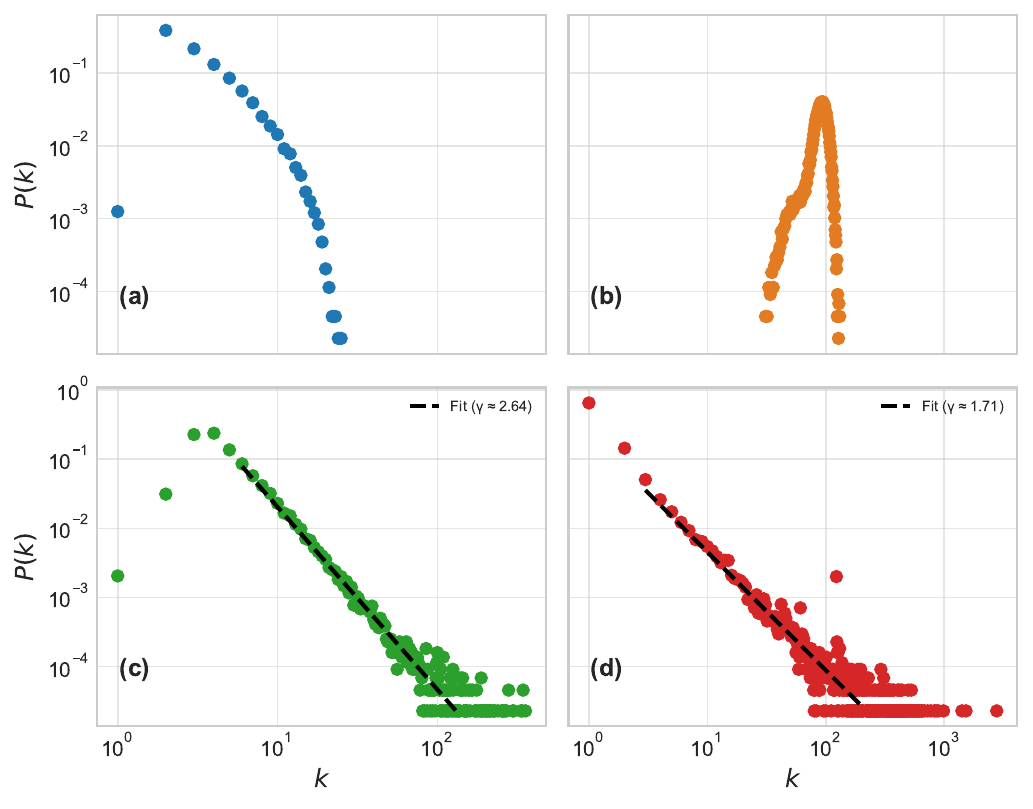}
        \captionsetup{labelformat=empty}
        \caption{Degree Distribution $P(k)$.}
    \end{subfigure}
    \hfill
    \begin{subfigure}[b]{0.32\textwidth}
        \centering
        \includegraphics[width=\textwidth]{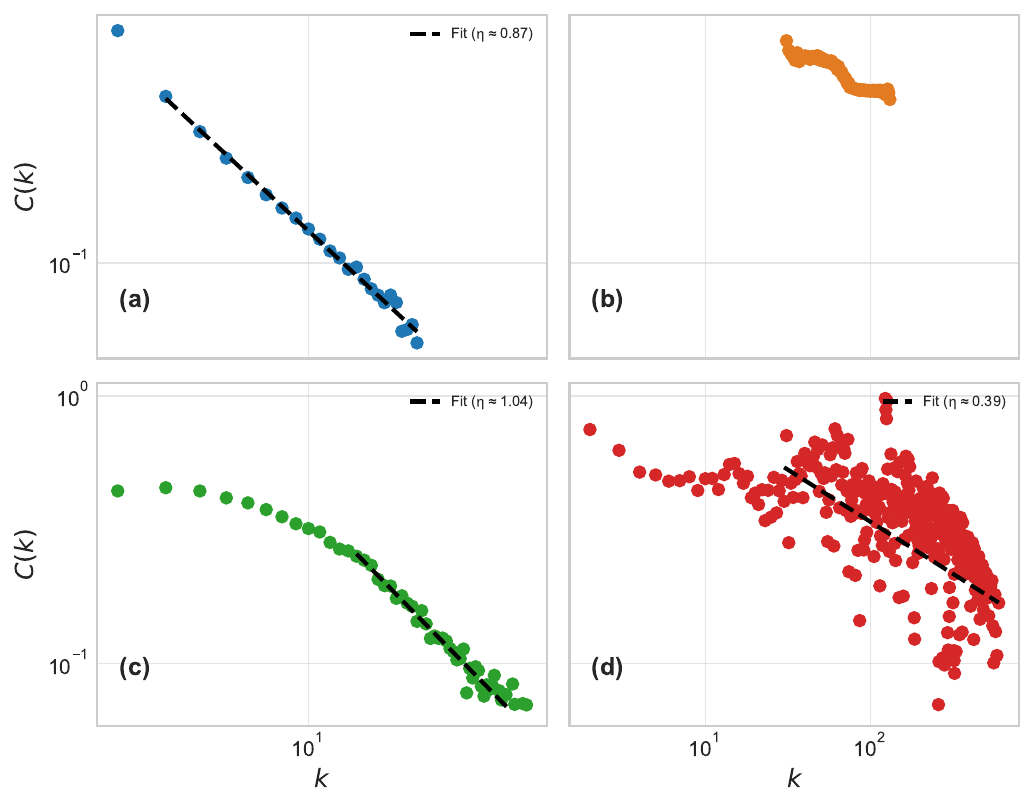}
        \captionsetup{labelformat=empty}
        \caption{Clustering Coefficient $C(k)$.}
    \end{subfigure}
    \hfill
    \begin{subfigure}[b]{0.32\textwidth}
        \centering
        \includegraphics[width=\textwidth]{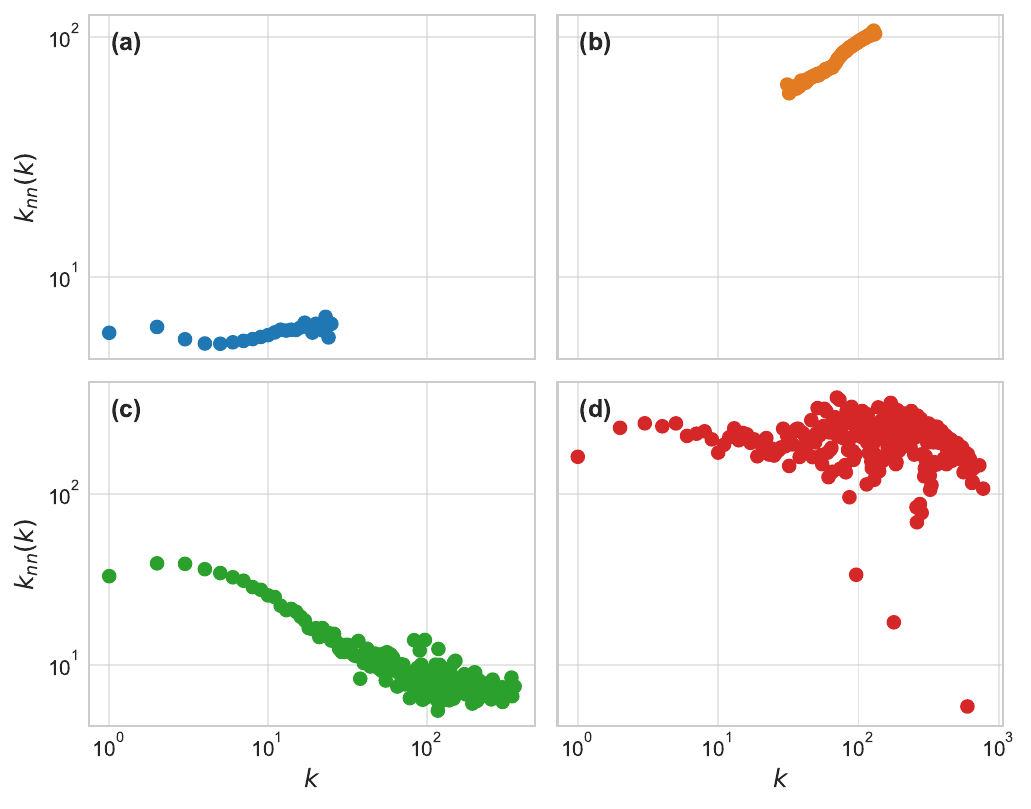}
        \captionsetup{labelformat=empty}
        \caption{K-Nearest Neighbors ($k_{nn}(k)$).}
    \end{subfigure}

    \caption{Analysis of the North America network. The top panel shows the map of geolocated nodes, and the bottom panels show the degree distribution $P(k)$, clustering coefficient $C(k)$, and K-Nearest Neighbors ($k_{nn}(k)$) for the different networks. In each subfigure we show the comparison of  (a) Brito-Soares , (b) Brito et al., (c) Brito-Rozenfeld, (d) Real Network.}
    \label{fig:northamerica}
\end{figure}

\begin{figure}[h!]
    \centering
    \begin{subfigure}[b]{0.9\textwidth}
        \centering
        \includegraphics[width=\textwidth]{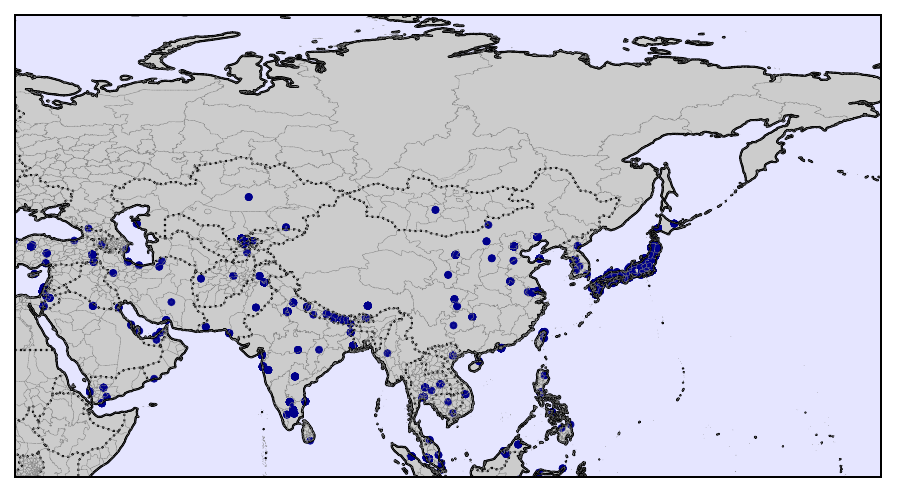}
        \captionsetup{labelformat=empty}
        \caption{Map of geolocated nodes.}
    \end{subfigure}

    \vskip 0.3cm

    \begin{subfigure}[b]{0.32\textwidth}
        \centering
        \includegraphics[width=\textwidth]{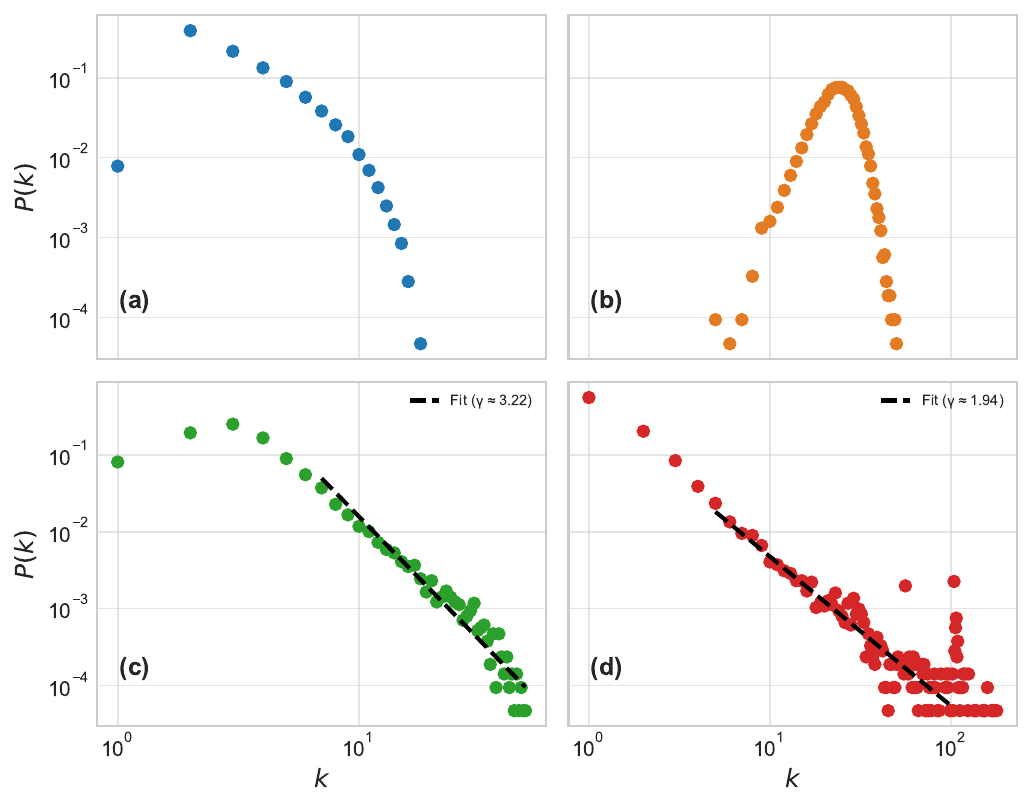}
        \captionsetup{labelformat=empty}
        \caption{Degree Distribution $P(k)$.}
    \end{subfigure}
    \hfill
    \begin{subfigure}[b]{0.32\textwidth}
        \centering
        \includegraphics[width=\textwidth]{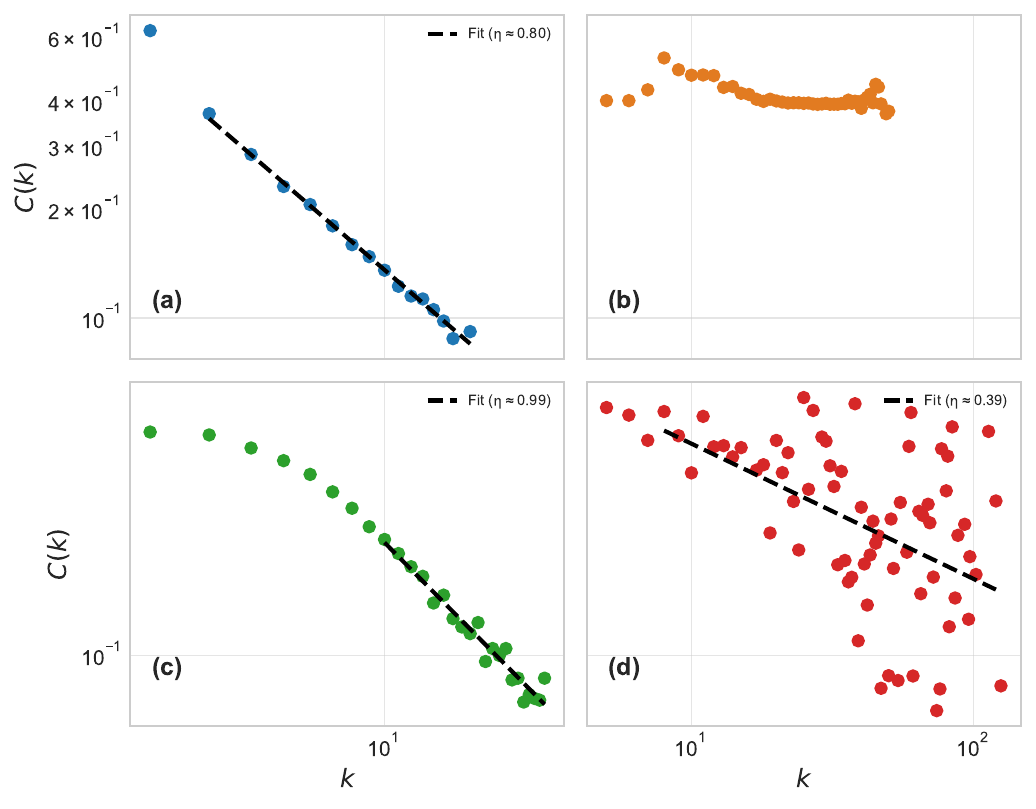}
        \captionsetup{labelformat=empty}
        \caption{Clustering Coefficient $C(k)$.}
    \end{subfigure}
    \hfill
    \begin{subfigure}[b]{0.32\textwidth}
        \centering
        \includegraphics[width=\textwidth]{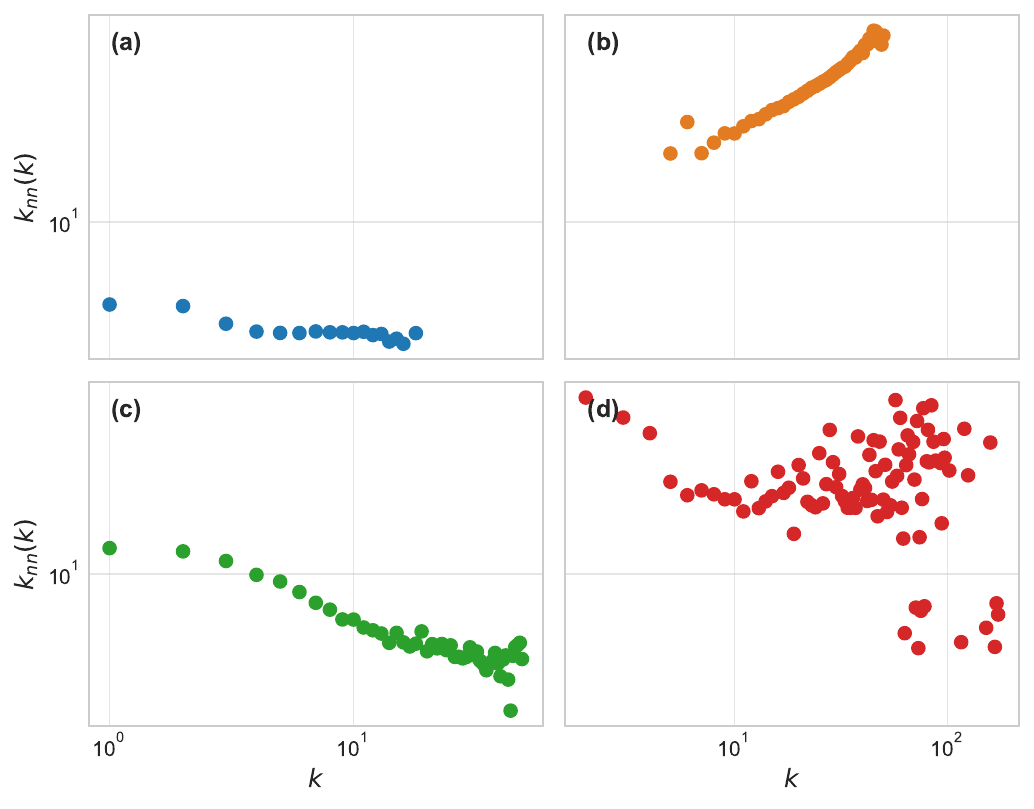}
        \captionsetup{labelformat=empty}
        \caption{K-Nearest Neighbors ($k_{nn}(k)$).}
    \end{subfigure}

    \caption{Analysis of the Asia network. The top panel shows the map of geolocated nodes, and the bottom panels show the degree distribution $P(k)$, clustering coefficient $C(k)$, and K-Nearest Neighbors ($k_{nn}(k)$) for the different networks. In each subfigure we show the comparison of  (a) Brito-Soares , (b) Brito et al., (c) Brito-Rozenfeld, (d) Real Network.}
    \label{fig:northamerica}
\end{figure}

\begin{figure}[h!]
    \centering
    \begin{subfigure}[b]{0.9\textwidth}
        \centering
        \includegraphics[width=\textwidth]{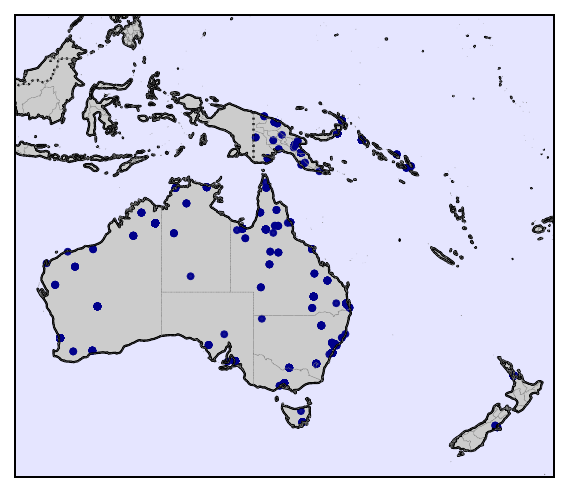}
        \captionsetup{labelformat=empty}
        \caption{Map of geolocated nodes.}
    \end{subfigure}

    \vskip 0.3cm

    \begin{subfigure}[b]{0.32\textwidth}
        \centering
        \includegraphics[width=\textwidth]{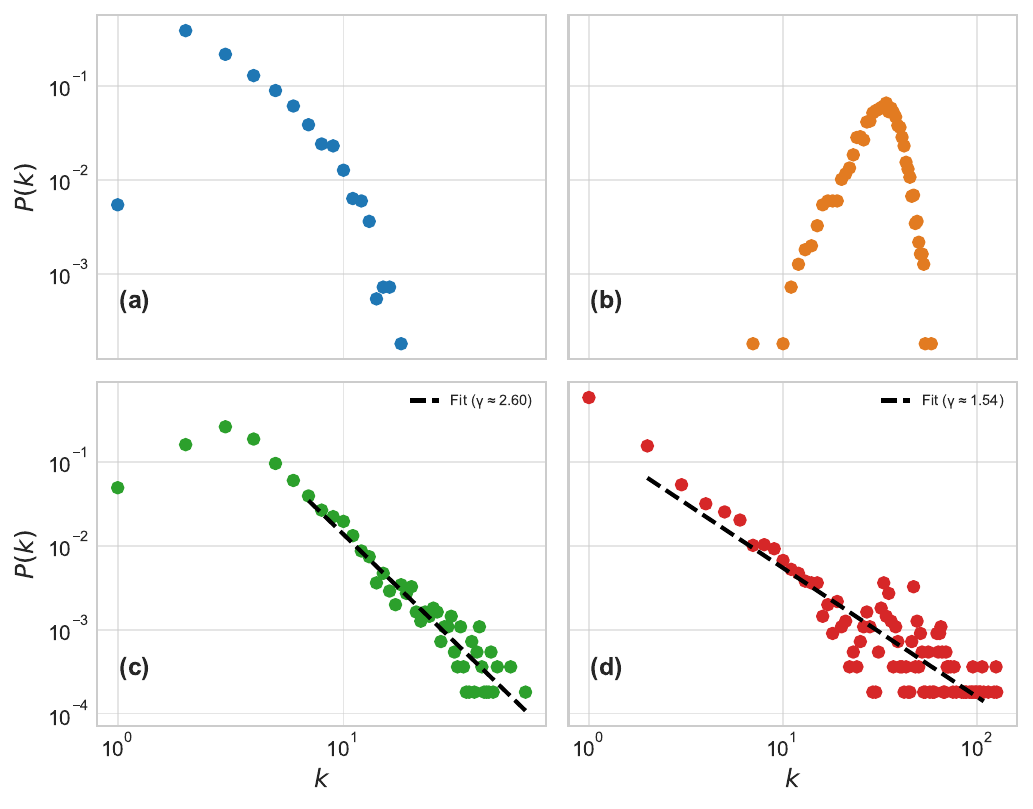}
        \captionsetup{labelformat=empty}
        \caption{Degree Distribution $P(k)$.}
    \end{subfigure}
    \hfill
    \begin{subfigure}[b]{0.32\textwidth}
        \centering
        \includegraphics[width=\textwidth]{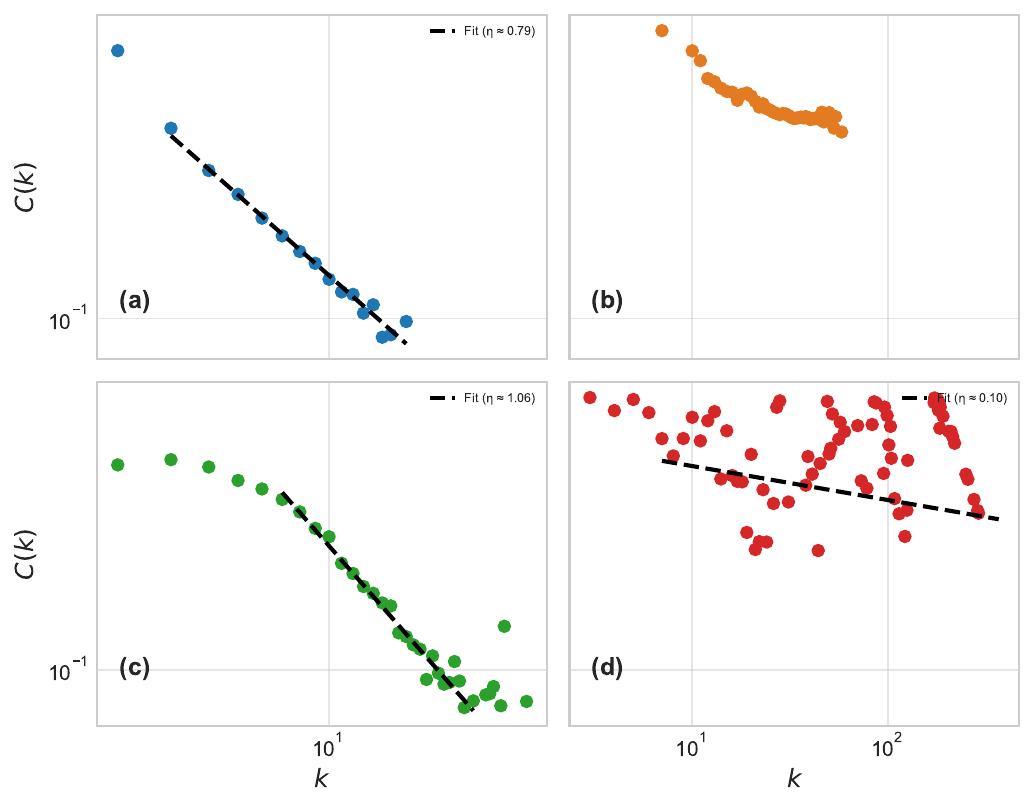}
        \captionsetup{labelformat=empty}
        \caption{Clustering Coefficient $C(k)$.}
    \end{subfigure}
    \hfill
    \begin{subfigure}[b]{0.32\textwidth}
        \centering
        \includegraphics[width=\textwidth]{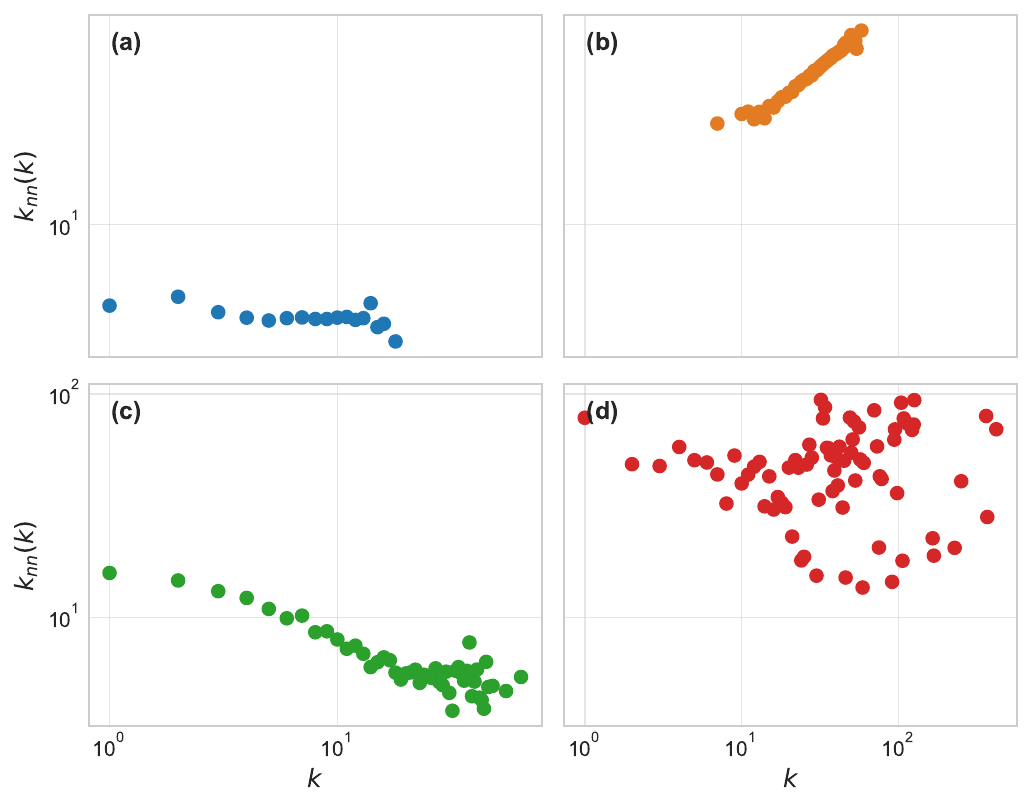}
        \captionsetup{labelformat=empty}
        \caption{K-Nearest Neighbors ($k_{nn}(k)$).}
    \end{subfigure}

    \caption{Analysis of the Oceania network. The top panel shows the map of geolocated nodes, and the bottom panels show the degree distribution $P(k)$, clustering coefficient $C(k)$, and K-Nearest Neighbors ($k_{nn}(k)$) for the different networks. In each subfigure we show the comparison of  (a) Brito-Soares , (b) Brito et al., (c) Brito-Rozenfeld, (d) Real Network.}
    \label{fig:northamerica}
\end{figure}

\end{document}